\documentclass[twocolumn,reprint,showpacs,preprintnumbers,amsmath,amssymb,superscriptaddress]{revtex4-1}
\usepackage{graphicx}
\usepackage{dcolumn}
\usepackage{bm}
\usepackage{graphicx,color}
\usepackage{subfigure}
\usepackage{physics}
\usepackage{tikz}
\usetikzlibrary{automata,positioning}
\usepackage{epstopdf}

\begin{document}

\title{Classical information driven quantum dot thermal machines}

\author{Abhin Shah}
\affiliation{Department of Electrical Engineering, Indian Institute of Technology Bombay, Powai, Mumbai-400076, India.\\}
\author{Sai Vinjanampathy}
 \affiliation{Department of Physics, Indian Institute of Technology Bombay, Powai, Mumbai-400076, India.\\}
 \affiliation{Centre for Quantum Technologies, National University of Singapore, 3 Science Drive 2, Singapore 117543.\\}
\author{Bhaskaran Muralidharan }
\affiliation{Department of Electrical Engineering, Indian Institute of Technology Bombay, Powai, Mumbai-400076, India.\\}



\date{\today}

\begin{abstract}
We analyze the transient response of quantum dot thermal machines that can be driven by hyperfine interaction acting as a source of classical information.\ Our setup comprises a quantum dot coupled to two contacts that drive heat flow while coupled to a nuclear spin bath.\ The quantum dot thermal machines operate both as batteries and as engines, depending on the parameter range. The electrons in the quantum dot interact with the nuclear spins via hyperfine spin-flip processes as typically seen in solid state systems such as GaAs quantum dots. The hyperfine interaction in such systems, which is often treated as a deterrent for quantum information processing, can favorably be regarded as a driving agent for classical information flow into a heat engine setup.\ We relate this information flow to Landauer's erasure of the nuclear spin bath, leading to a battery operation. We further demonstrate that the setup can perform as a transient power source even under a voltage bias across the dot. Focusing on the transient thermoelectric operation, our analysis clearly indicates the role of Landauer's erasure to deliver a higher output power than a conventional quantum dot thermoelectric setup and an efficiency greater than that of an identical Carnot cycle in steady state, which is consistent with recently proposed bounds on efficiency for systems subject to a feedback controller. The role of nuclear spin relaxation processes on these aspects is also studied. Finally, we introduce the Coulomb interaction in the dot and analyze the transient thermoelectric response of the system. Our results elaborate on the effective use of somewhat undesirable scattering processes as a non-equilibrium source of Shannon information flow in thermal machines and the possibilities that may arise from the use of a quantum information source.
\end{abstract}
\maketitle
\section{\label{sec:level1}Introduction}
Fundamental thermoelectric transport studies aimed at probing the physics of heat flow in the nanoscale \cite{Mahan1996,Andreev2001,Humphrey2002,Kubala2006,Kubala2008,nak,Kim2014,Reddy2007,jordan1,jordan2,Agarwal2014,whitney, Basky_Grifoni,Zimb2016} have been very actively pursued in recent times. In this context, the quantum dot thermoelectric setup \cite{Leijnse,Humphrey2002,Basky_Grifoni,nak,Basky_Milena,jordan3,sothmann,bitan,Sanchez} is an ideal test bed and a minimal model to understand advanced concepts related to the microscopics of heat flow.\ In recent times, there is also considerable interest in understanding the intricate connection between information and thermodynamics \cite{Datta_Demon,Abreu_Seifert,Bauer_Seifert,David_Seifert,Seifert2014,Esposito_thermo,Esposito_Demon,Averin}. It has also been shown \cite{Datta_Demon,Esposito_Demon,Jarzynski,Wieck,Esposito_thermo} that ``demon" assisted transport setups can be devised to work as a battery.\ These setups typically involve the active channel being coupled to ancillary systems that act as demons \cite{Datta_Demon}.\ The action of the demon ancilla may also be thought of as a feedback controller on the channel, which is the quantum dot in our case.\ In such cases, important bounds on the thermodynamic efficiencies have been proposed \cite{Esposito_thermo,Sagawa2008,Sagawa2009,Sagawa2010,Sagawa2014}. \\
\indent At the same time, the action of the demon ancilla may also be thought of in terms of a flow in Shannon information into the active channel which may be viewed as the reverse process of Landauer's erasure \cite{Datta_Demon,Averin}. While there has been a lot of recent attention to improve quantum thermal machines using quantum coherence and entanglement \cite{goold2016role,vinjanampathy2016quantum}, somewhat less attention has been given to improving \textit{quantum} thermal machines with classical information. In this manuscript, we analyze in detail such a quantum thermal machine that features a hyperfine mediated quantum dot setup and demonstrate an enhancement in the performance of quantum thermal machines driven by classical information.\\
\begin{figure*}[]
\begin{center}
\subfigure[]{\includegraphics[width=0.5\textwidth, height=0.25\textwidth]{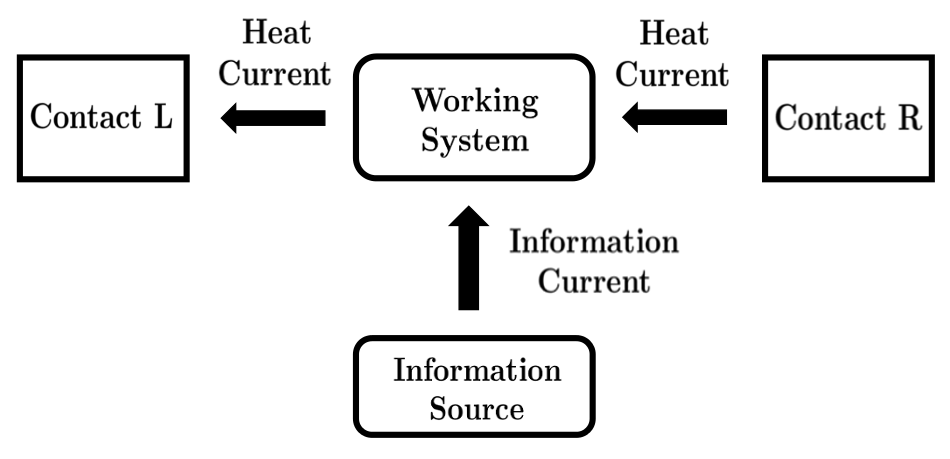}\label{1a}}
\quad
\subfigure[]{\includegraphics[width=0.55\textwidth, height=0.25\textwidth]{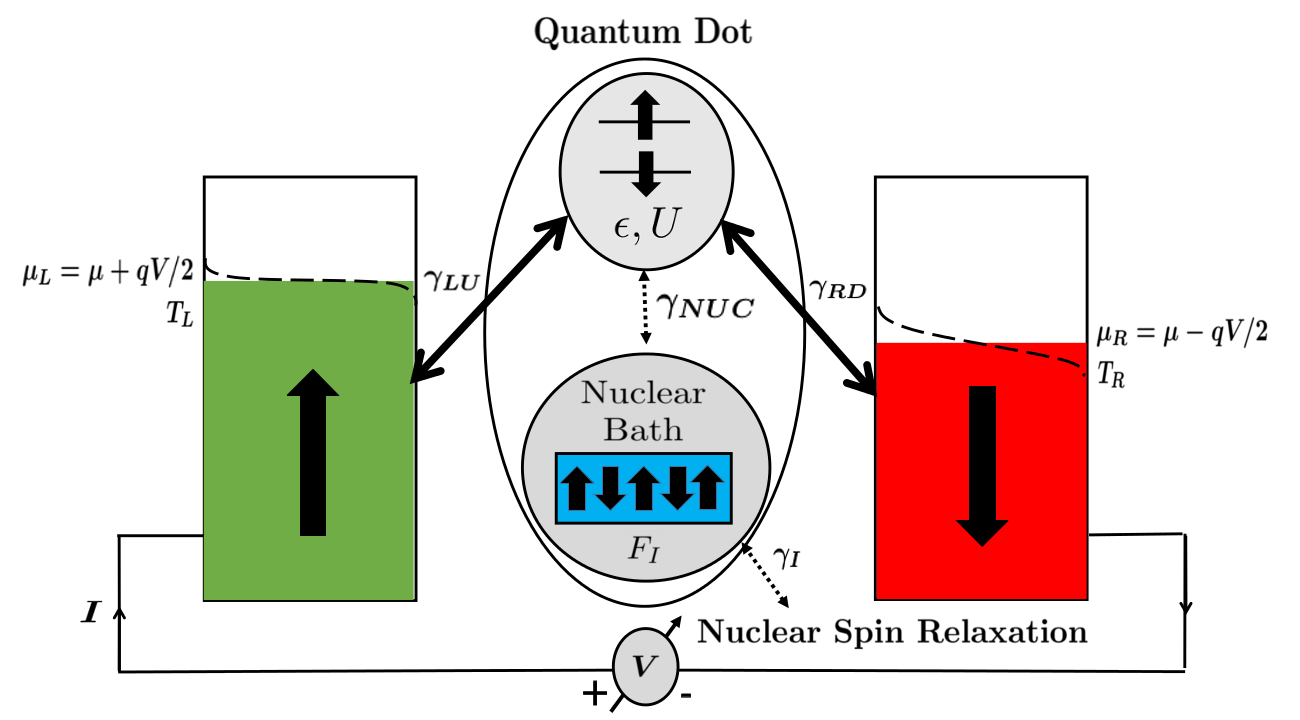}\label{1b}}
\quad
\subfigure[]{\includegraphics[width=0.3\textwidth, height=0.25\textwidth]{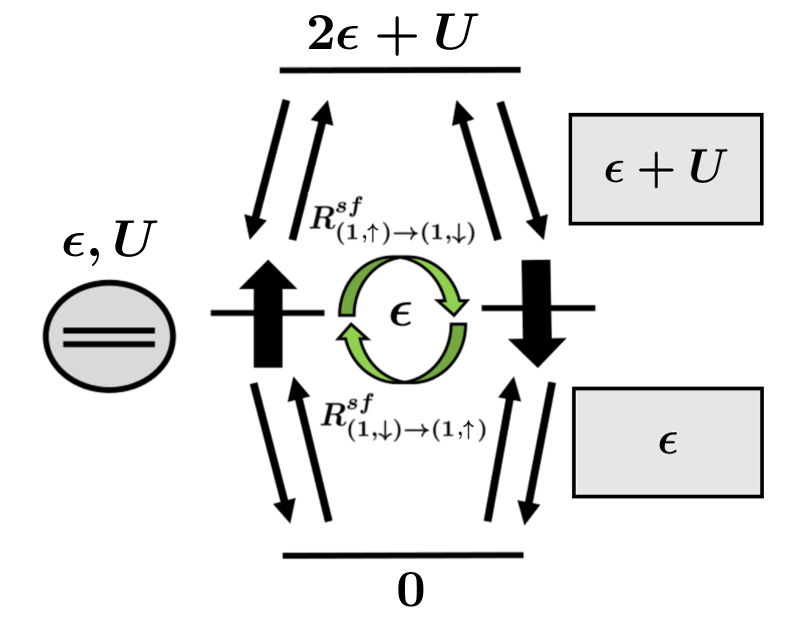}\label{1c}}
\end{center}
\caption{ Schematic of the information-driven quantum dot thermal machine: (a) The out-of-equilibrium information source drives the working system with a supply of information current in addition to the regular heat engine operation. (b) A physical realization of such a paradigm is described by single interacting quantum dot setup with spin-degenerate levels representing the working system. The dot is coupled to the left and right contacts with electronic rates $\gamma_{LU}$ and $\gamma_{RD}$ respectively. The dot is also coupled to a nuclear bath (information source) with spin-flip rate $\gamma_{nuc}$.  The nuclear bath can relax its spin depending on the nuclear spin relaxation constant $\gamma_{I}$. This setup can work like a battery, when $\Delta\mu = \mu_L - \mu_R = 0 $ and $\Delta T = T_R - T_L = 0$, as well as a heat engine when $\Delta\mu \neq 0 $ and $\Delta T \neq 0$.  (c) State transition diagram in the quantum dot electronic Fock space. Electronic transitions take place between states with electron numbers differing by $\pm$ 1 due to the contacts and between the spin degenerate levels in the one electron subspace via electron spin-flip transitions accompanied by nuclear spin-flop transitions.}
\end{figure*}
\indent A schematic of our thermal machine is depicted in Fig.~\ref{1a}.\ The setup we analyze comprises a quantum dot coupled to two contacts that drive heat flow, while coupled to a nuclear spin bath as schematized in Fig.~\ref{1b}.\ The electrons in the quantum dot interact with the nuclear spins via hyperfine spin-flip processes, which, in turn, acts as the driving agent for classical information flow into the heat engine setup.\ There has been sufficient theoretical and experimental research in the area of nuclear spintronics concerning the manipulation of nuclear spins by means of hyperfine interaction between the host nuclei and the itinerant electrons in the quantum transport setup \cite{Wieck,Hanson2007,Levitov1,Levitov2,Levitov3,Basky_buddhi,Aniket}, specifically with the aim of controlling the undesired effects of hyperfine interaction on the electronic qubit system. This specifically involves the study of dynamic nuclear polarization by transferring spin polarization from electrons to the nuclear spin system \cite{Hanson2007}.\ The so-called detrimental hyperfine spin-flip processes, we demonstrate, sets the stage to develop information driven quantum dot thermal machines, specifically using the well developed GaAs setups. We demonstrate that the act of nuclear spin assisted spin-flip scattering, can be cast as a flow of classical information source.\ The polarization of the nucleus serves as the information content, and the rate of change of Shannon entropy of the bath or the erasure rate is the information current. \\ 
We simulate the fast dynamics of electronic transport self-consistently with the slow dynamics of the nuclei \cite{Jesus,Basky_buddhi} and analyze the characteristics of the thermal machine.\ We show that (a) even in the absence of a voltage or a temperature gradient, the reverse Landauer's erasure process via hyperfine mediated spin-flips results in a non-zero electronic current flowing through the setup under transient conditions, thus functioning as a battery whose discharge characteristics are strongly correlated with the nuclear spin relaxation processes (b) even with a voltage bias across the dot, the system can drive a current in the external circuit and can thus be used as a power source to perform useful work, and (c) in the heat engine case, where both temperature gradient and voltage are impressed, the setup performance under transient conditions well exceeds that of steady state, in terms of the open circuit (Seebeck) voltage and efficiency.\ We note that under steady state conditions the performance of this setup compares favorably with other schemes \cite{Leijnse,Humphrey2002,Basky_Grifoni,nak,Basky_Milena,jordan3,sothmann,bitan,Sanchez}.\ Finally, we introduce the Coulomb interaction parameter and analyze the thermoelectric response of the system. \\

\indent This manuscript is organized as follows: Sec. II elucidates the generic information driven heat engine and the transport formulation. In Sec. IIIA we analyze the discharging characteristics of the system without any voltage or temperature bias across the contacts. In Sec. IIIB we illustrate the transient properties of the system with a potential difference between the contacts. In Sec IIIC. we study the thermoelectric performance of the dot with left contact maintained at a higher potential, and the right contact maintained at a higher temperature.\ Sec IV summarizes the main results of this work.
\section{\label{sec:level2}Physics and Formulation}
The system comprises a single orbital Anderson-impurity-type quantum dot described by the following one-site Hubbard Hamiltonian (Fig.~\ref{1b}):
\begin{equation}
\begin{split}
\hat{H}_{S}= \epsilon (\hat{n}_{\uparrow} + \hat{n}_{\downarrow})+U \hat{n}_{\uparrow}\hat{n}_{\downarrow},
\end{split}
\end{equation}
where $\epsilon$ represents the orbital energy, $ \hat{n}_{\uparrow (\downarrow)}$ is the occupation number operator of an electron with spin $\uparrow$($\downarrow$), and U is the Coulomb interaction between electrons of opposite spins occupying the same orbital. We consider only spin-degenerate levels such that $\epsilon_\uparrow = \epsilon_\downarrow = \epsilon$. This system is weakly coupled to two ferromagnetic contacts that are fully polarized, the left which is spin-polarized in the up direction and the right which is spin-polarized in the down direction.\ The coupling is described by the typical tunneling Hamiltonian between the contact states and the device states as described by many related works \cite{Basky_Milena,Basky_Grifoni,bitan,Leijnse}. The contacts are denoted as $ L $ (left)  and $ R $ (right), each of which is characterized by a temperature $T_{L(R)}$ and an electrochemical potential $\mu_{L(R)}$. \\
\indent Additionally, the dot is coupled with a nuclear bath \cite{Basky_buddhi} whose nuclei interact with the itinerant electrons in the quantum dot via the Fermi-contact hyperfine interaction \cite{Basky_buddhi,Aniket} which is given by $\hat{H}_{HF} = \sum_{k=1}^{N_{nuc}}J_k \hat{{\bf{I}}}_k\cdot \hat{{\bf{S}}}$, where $\hat{{\bf{S}}}$ is the electronic spin operator, $\hat{{\bf{I}}}_k$ is the nuclear spin operator and $J_k = J_{eff}\nu_0|\psi_k|^2$ is the hyperfine interaction parameter of an individual nucleus treated as a point particle.\ Here, $J_{eff}$ is a material specific hyperfine coupling parameter, $\nu_0$ is the volume of the unit cell of the nucleus, $\psi_k$ the electronic wavefunction at the nucleus site $k$, and $N_{nuc}$ is the number of nuclei in the nuclear bath. This Hamiltonian can be expanded as
\begin{equation}\label{ham}
\hat{H}_{HF} = \sum_{k=1}^{N_{nuc}}J_k \hat{I}_k^z \hat{S}_z + \frac{1}{2}\sum_{k=1}^{N_{nuc}}J_k( \hat{I}^-_{k} \hat{S}^+ + \hat{I}^+_k \hat{S}^-),
\end{equation}
whose second term is the {{spin-flip}} part, denoted by $\hat{H}_{sf}$. Under mean field approximation, the first term may be treated as an effective magnetic field on the electrons within the electronic Hamiltonian and is referred to as the Overhauser field \cite{Basky_buddhi, Aniket}, which we neglect in this work. The reason for this is that the Overhauser field depends on the value of the hyperfine exchange parameter, which is usually very small, and the polarization of the nuclei. Its effect on transport is to split the degenerate levels and also cause some bistable effects. These effects are of higher order and do not contribute much to the basic ideas conveyed here. It is the $\hat{H}_{sf}$ term that triggers the spin-flip processes which in our formalism are described via the evaluation of spin-flip rates to be described shortly. Under the assumption of weak hyperfine coupling, which is typically the case in typical systems \cite{Hanson2007}, the spin-flip rates can be evaluated via the Fermi's golden rule. The spin-flip rates will ultimately be related to the information current supplied from the nuclear bath to the electronic system. 
\subsubsection{Transport formulation}
We work in the sequential tunneling limit( $\hbar\gamma \ll k_BT$), where, $\gamma$ refers to a generic transition rate, which may include the tunnel transition, the spin-flip transition and other relaxation processes. In this limit, transport is described via rate equations \cite{Beenakker,Basky_Ghosh,Basky_Datta,Timm}. Furthermore, since we are assuming collinear leads, it suffices to work  in the diagonal subspace of the reduced density matrix of the system \cite{Koenig_1,Koenig_2,Brouw,Milena_noncoll,Basky_Milena}. Hence the electronic transport is described in terms of the occupation probabilities, $P_{0}^{0}$, $P^{1}_{\uparrow}$, $P^{1}_{\downarrow}$, $P_{0}^{2}$, of each of the Fock space states shown in  Fig.~\ref{1c} with total energies $0$, $\epsilon$, $\epsilon$, and $2\epsilon + U$. The many-body master equation approach incorporates spin-flip transition rates $R^{sf}_{(1, \uparrow) \rightarrow (1, \downarrow)}$ and $R^{sf}_{(1, \downarrow) \rightarrow (1, \uparrow)}$ between the states $| 1,\uparrow \rangle$ and  $| 1,\downarrow \rangle$ having different spin symmetries with the same number of electrons. The tunneling transition rates for all possible transitions are given as,
\begin{equation}
\begin{split}
R_{(0,0)\rightarrow(1,\uparrow)} &= \gamma_{LU} \cdot  f\big(\frac{\epsilon-\mu_{L}}{k_{B} T_{L}}\big)  \\
R_{(0,0)\rightarrow(1,\downarrow)} &=  \gamma_{RD} \cdot  f\big(\frac{\epsilon-\mu_{R}}{k_{B} T_{R}}\big)  \\
R_{(1,\uparrow)\rightarrow(2,0)} &=  \gamma_{RD} \cdot  f\big(\frac{\epsilon + U -\mu_{R}}{k_{B} T_{R}}\big)  \\
R_{(1,\downarrow)\rightarrow(2,0)} &=  \gamma_{LU} \cdot  f\big(\frac{\epsilon + U -\mu_{L}}{k_{B} T_{L}}\big)  \\
R_{(1,\uparrow)\rightarrow(0,0)}&= \gamma_{LU} \cdot  \big[1- f\big(\frac{\epsilon-\mu_{L}}{k_{B} T_{L}}\big) \big]  \\
R_{(1,\downarrow)\rightarrow(0,0)}&=  \gamma_{RD} \cdot \big[1 - f\big(\frac{\epsilon-\mu_{R}}{k_{B} T_{R}}\big)  \big]\\
R_{(2,0)\rightarrow(1,\uparrow)} &= \gamma_{RD} \cdot  \big[1 - f\big(\frac{\epsilon + U -\mu_{R}}{k_{B} T_{R}}\big)  \big]\\
R_{(2,0)\rightarrow(1,\downarrow)} &= \gamma_{LU} \cdot \big[1 -  f\big(\frac{\epsilon + U -\mu_{L}}{k_{B} T_{L}}\big)\big]
\end{split}
\end{equation}
where $ f $ is the Fermi-Dirac distribution function. In the anti-parallel ferromagnetic configuration that we consider, we take $\gamma_{LD} = \gamma_{RU} = 0 $ and thus we effectively have $\gamma_{LU}$ and $\gamma_{RD}$ to be the electronic coupling rates for the left contact and the right contact respectively. Note that this restricts the association of inflow and outflow of up (down) spin electron with only the left (right) contact. 
\\ \indent We define the nuclear spin polarization \cite{Basky_buddhi} $F_I=\frac{N_\Uparrow - N_\Downarrow}{N_{nuc}} $ where  $N_\Uparrow $ and $N_\Downarrow$ are the number of up-spin and down-spin nuclei respectively. Using $ N_{nuc} = N_\Uparrow + N_\Downarrow $, we get
\begin{equation}
\begin{split}
P_k^{\Downarrow} = N_\Downarrow/N_{nuc} = (1-F_I)/2 \\
P_k^{\Uparrow} = N_\Uparrow/N_{nuc} = (1+F_I)/2
\end{split}
\end{equation}
where $P_k^{\Uparrow(\Downarrow)}$ is the probability that the $k^{th}$ nucleus is in the up (down) spin state.\\ 
\indent The spin-flip rates \cite{Basky_buddhi} for the electrons are evaluated via the Fermi's golden rule in which the rate of transition from an initial state in the $|\uparrow \rangle$ to a final state $|\downarrow \rangle $ in the electronic Fock space is given by 
\begin{equation}
\begin{split}
R^{sf}_{(1 , \uparrow) \rightarrow (1, \downarrow)} = \frac{\pi}{2\hbar}\frac{|J_{eff}|^2}{N_{nuc}} \rho(E)\frac{1-F_I}{2},
\end{split}
\end{equation}
and similarly,
\begin{equation}
\begin{split}
\\R^{sf}_{(1, \downarrow) \rightarrow (1, \uparrow)} = \frac{\pi}{2\hbar}\frac{|J_{eff}|^2}{N_{nuc}} \rho(E)\frac{1+F_I}{2}
\end{split}
\end{equation}
Here, $\rho (E = \epsilon_\uparrow - \epsilon_\downarrow) \rightarrow \frac{\eta}{( \epsilon_\uparrow - \epsilon_\downarrow)^2+ \eta^2} = \frac{1}{\eta}$ represents the Lorentzian density of states associated with a spin-flip transition and $\eta$ is the lifetime damping parameter (assumed to be of the order of 0.1$\mu$eV). The value of $\eta$  is of the order of the lifetime of the blocked state, which in this case could be the up spin or the down spin state, depending on which contact initiates the charge flow. With various rates defined above, the master equation for the probabilities $P^{N}_{i}$, defined by the size of the electronic Fock space, reads as:
\begin{equation}
\begin{split}
\frac{d P^{0}_{0}}{dt}= - R_{(0,0)\rightarrow(1,\uparrow)}P^{0}_{0}   - R_{(0,0)\rightarrow(1,\downarrow)}P^{0}_{0}\\ + R_{(1,\uparrow)\rightarrow(0,0)}P^{1}_{\uparrow} + R_{(1,\downarrow)\rightarrow(0,0)}P^{1}_{\downarrow}\\
\frac{d P^{1}_{\uparrow}}{dt}= - R_{(1,\uparrow)\rightarrow(0,0)}P^{1}_{\uparrow}   - R_{(1,\uparrow)\rightarrow(2,0)}P^{1}_{\uparrow}\\ + R_{(2,0)\rightarrow(1,\uparrow)}P^{2}_{0} + R_{(0,0)\rightarrow(1,\uparrow)}P^{0}_{0}\\ + R^{sf}_{(1,\downarrow)\rightarrow(1,\uparrow)}P^{1}_{\downarrow}- R^{sf}_{(1,\uparrow)\rightarrow(1,\downarrow)}P^{1}_{\uparrow} \\
\frac{d P^{1}_{\downarrow}}{dt}= - R_{(1,\downarrow)\rightarrow(0,0)}P^{1}_{\downarrow}   - R_{(1,\downarrow)\rightarrow(2,0)}P^{1}_{\downarrow}\\ + R_{(2,0)\rightarrow(1,\downarrow)}P^{2}_{0} + R_{(0,0)\rightarrow(1,\downarrow)}P^{0}_{0}\\ + R^{sf}_{(1,\uparrow)\rightarrow(1,\downarrow)}P^{1}_{\uparrow}- R^{sf}_{(1,\downarrow)\rightarrow(1,\uparrow)}P^{1}_{\downarrow}
\end{split}
\label{eq:elec_dy}
\end{equation}
along with the normalization equation $\sum_{N,i}P^{N}_{i} = 1 $. \\
\indent Having computed all the necessary electronic rates, we obtain the dynamics of the collective nuclear polarization $F_I $ from the individual master equations \cite{Basky_buddhi} for $N_\Uparrow $ and $N_\Downarrow$ as
\begin{equation}
\begin{split}
\frac{dF_I}{dt} = \gamma^{nuc} [(P^{1}_{\uparrow} - P^{1}_{\downarrow}) - (P^{1}_{\uparrow} + P^{1}_{\downarrow})F_I] - \gamma_{I} F_{I}
\end{split}
\label{eq:nuc_dy}
\end{equation}
where $\gamma_{nuc} = \frac{\pi}{2\hbar}\frac{|J_{eff}|^2}{N_{nuc}^2}\rho(E)$ with $N_{nuc} = 10^3$, $J_{eff} = 0.1\mu eV$ and $\hbar\gamma_I \sim 10^{-11}$ meV. $\gamma_I$ is a phenomenological nuclear spin relaxation constant. It represents how quickly the spins forget the direction in which they are oriented.\\
\indent Typically, timescales associated with nuclear spin relaxation are of the order of a few seconds which is very long compared to the electron-transport time scales\cite{Basky_buddhi}. Hence we can decouple the fast dynamics of electronic transport \eqref{eq:elec_dy} from the slow dynamics of the nuclei \eqref{eq:nuc_dy} and set $\frac{d P^{N}_{i}}{dt}=0$, and find the null space of the rate matrix to evaluate the transient occupation probabilities.  This is solved self consistently\cite{MacDonald,Basky_buddhi} with the slowly varying nuclear dynamics, which involves the evaluation of the spin-flip transition rates $R^{sf}$. Using the transient and steady state probabilities, we can get the expressions for the terminal electronic charge currents $I^\alpha$ and the electronic heat currents $I^{\alpha}_{Q}$, as
\begin{equation}
\begin{split}
I^{\alpha}=-q \bigg[R_{(0,0) \rightarrow(1,\uparrow)}^{\alpha}P^{0}_{0} + R_{(0,0) \rightarrow(1,\downarrow)}^{\alpha}P^{0}_{0}\\  -  R_{(1,\uparrow)\rightarrow(0,0)}^{\alpha}P^{1}_{\uparrow}  - R_{(1,\downarrow)\rightarrow(0,0)}^{\alpha}P^{1}_{\downarrow}\\ +  R_{(1,\uparrow)\rightarrow(2,0)}^{\alpha}P^{1}_{\uparrow}  + R_{(1,\downarrow)\rightarrow(2,0)}^{\alpha}P^{1}_{\downarrow}\\ - R_{(2,0) \rightarrow(1,\uparrow)}^{\alpha}P^{2}_{0} - R_{(2,0) \rightarrow(1,\downarrow)}^{\alpha}P^{2}_{0}
\bigg]
\end{split}
\end{equation}
\begin{equation}
\begin{split}
I^{\alpha}_Q=(\epsilon - \mu_\alpha)R_{(0,0) \rightarrow(1,\uparrow)}^{\alpha}P^{0}_{0} + (\epsilon - \mu_\alpha + U)R_{(1,\uparrow)\rightarrow(2,0)}^{\alpha}P^{1}_{\uparrow} \\
 (\epsilon - \mu_\alpha)R_{(0,0) \rightarrow(1,\downarrow)}^{\alpha}P^{0}_{0}  + (\epsilon - \mu_\alpha + U)R_{(1,\downarrow)\rightarrow(2,0)}^{\alpha}P^{1}_{\downarrow}\\
 - (\epsilon - \mu_\alpha)R_{(1,\uparrow)\rightarrow(0,0)}^{\alpha}P^{1}_{\uparrow}  - (\epsilon - \mu_\alpha + U)R_{(2,0) \rightarrow(1,\uparrow)}^{\alpha}P^{2}_{0}\\
 - (\epsilon - \mu_\alpha)R_{(1,\downarrow)\rightarrow(0,0)}^{\alpha}P^{1}_{\downarrow} -  (\epsilon - \mu_\alpha + U)R_{(2,0) \rightarrow(1,\downarrow)}^{\alpha}P^{2}_{0}.
\end{split}
\end{equation}
The charge or electronic heat currents associated with contacts $\alpha=L(R)$ involve only the rates associated with the respective contact.\\
\indent The steady state nuclear spin polarization is given by setting $\frac{dF_I}{dt} =0 $ in \eqref{eq:nuc_dy}.
\begin{equation}
\begin{split}
F_I = \frac{\gamma^{nuc}(P^{1}_{\uparrow} - P^{1}_{\downarrow})}{\gamma_{I} + \gamma^{nuc}(P^{1}_{\uparrow} + P^{1}_{\downarrow})}
\label{eq:FI}
\end{split}
\end{equation}
The time dependence of the nuclear bath's Shannon entropy is given by
\begin{equation}
\begin{split}
 S(t) = -k_{B} \sum_{k} \bigg[ P_k^{\Uparrow}(t) \ln P_k^{\Uparrow}(t) + P_k^{\Downarrow}(t) \ln P_k^{\Downarrow}(t) \bigg]
\end{split}
\end{equation}
Now, we define the information current $(I_F)$ as the rate of change of Shannon entropy of the bath\cite{Esposito_thermo} $(\frac{dS_I}{dt})$, from which we can evaluate it as 
\begin{equation}
\begin{split}
I_F =X_I\left\{ \gamma^{nuc} [(P^{1}_{\uparrow} - P^{1}_{\downarrow}) - (P^{1}_{\uparrow} + P^{1}_{\downarrow})F_I] - \gamma_{I} F_{I}\right\}.
\end{split}
\end{equation}
It is interesting to note that the term in the curly bracket of the above equation relates to the so-called ``spin-flip'' current \cite{Datta_Demon}, and represents the virtual flow between the up-spin and down-spin states within the quantum dot. The term $X_I:=(k_{B}/2)\ln[(1+F_I)/(1-F_I)]$, represents a factor that is related to the rate of change of Shannon information. While physical quantities such as the polarization $F_I$ vary with a characteristic time-constant, the term $X_I$ modulates the characteristic time constant of the information current. Its implication on the current system will be obvious in the upcoming sections, where the information current is seen to decay at a much faster rate in comparison to the nuclear polarization, charge currents and other physical quantities, distinctly pointing toward the logarithmic dependence of information.
\subsubsection{Calculation of power and efficiency}
The instantaneous electrical power generated in the circuit is given by $P=-I^{\alpha}\times V$. In general, one can say that energy conversion occurs when $P > 0$.  In the thermoelectric case, this requires heat current from the hot contact $I^{R}_Q > 0 $ when a thermal bias is applied across the contacts, $T_R$, at the hot contact $R$, and $T_L$, at the cold contact $L$. At a voltage $V_S$, called the Seebeck or open circuit voltage, the back flow current completely cancels the charge current setup by the temperature gradient and the flow of information current. We will see that the open circuit voltage varies with time depending on the flow of information current. The setup thus functions as a heat engine in the voltage range $[0,V_S]$. The thermoelectric efficiency is expressed as
\begin{equation}
\eta=\frac{P}{I_Q^{R}}.
\label{eq:eff}
\end{equation}
We study the performance of the dot up to a voltage bias of $10$ meV. In all our calculations, we assume that half of the applied voltage drops across each tunnel barrier as a result of equal capacitive coupling to the two contacts. We have $T_L = 10 K$ and $T_R = T_L/(1-n_C)$ where $n_C$ is the Carnot efficiency. With no voltage bias, we have $\mu_L = \mu_R  = \mu.$ We have set the orbital energy $\epsilon - \mu= 2k_{B}T_R$. In the rest of the manuscript, we denote $I^R$ by $I$ and it is defined to be positive when the current flows from the left contact via the dot to the right contact. We consider the Coulomb interaction parameter to be zero in most of our analysis. Towards the end of Section~\ref{sec:level3} we introduce the Coulomb interaction parameter and re-analyze the transient thermoelectric response of the system.
\section{\label{sec:level3}Results}
\subsection{Discharging of an information-driven battery}
\begin{center}
\begin{figure}[!htb]
\subfigure[]{\includegraphics[height=0.225\textwidth,width=0.225\textwidth]{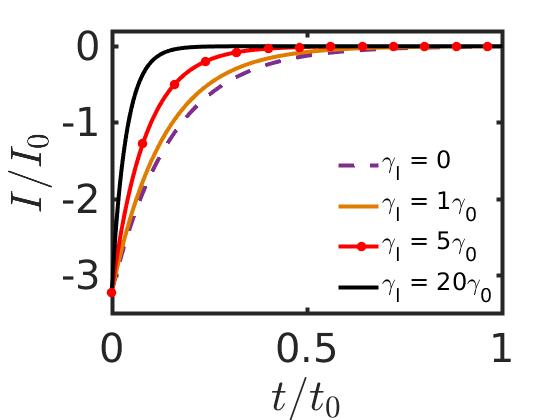}\label{2a}}
\quad
\subfigure[]{\includegraphics[height=0.225\textwidth,width=0.225\textwidth]{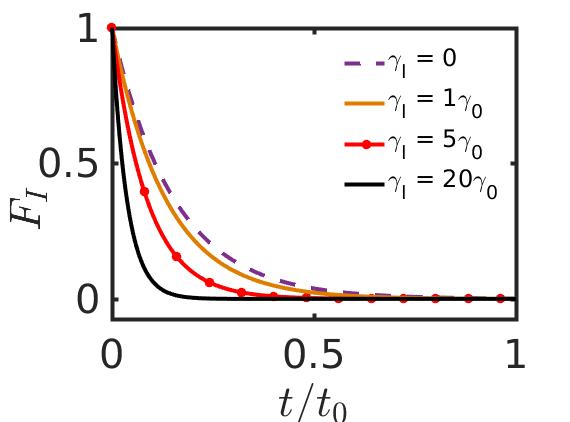}\label{2b}}
\quad
\subfigure[]{\includegraphics[height=0.225\textwidth,width=0.225\textwidth]{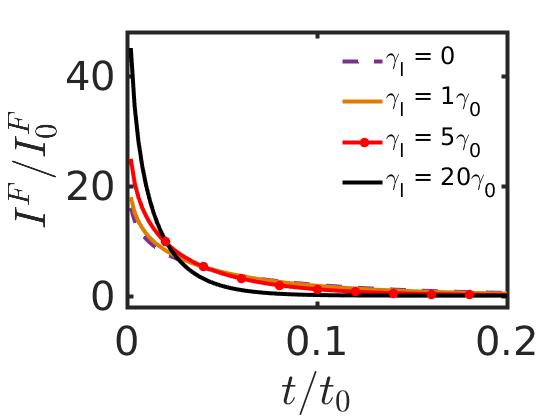}\label{2c}}
\quad		
\caption{Discharging of the information-driven battery with $\Delta T = 0 $ and $\Delta \mu = 0$. (a) The decay of the electronic current as a function of time as  $\gamma_I$ is varied. (b)  Variation in the nuclear spin polarization as a function of time as  $\gamma_I$ is varied. The current characteristics closely follow that of the nuclear spin polarization, and they both eventually decay to zero at the same time. (c) The information current as a function of time as  $\gamma_I$ is varied. The decay in $I_F$ is much faster compared to decay in $I$ and $F_I$ but the rate of decay of all three increases with increase in $\gamma_I$ }
\end{figure}
\end{center}

We first elaborate on the discharging characteristics of the information-driven battery. Surprisingly, this setup can perform as a battery without any temperature or voltage gradient, owing to the flow of information current. Without the nuclear bath, there can be no current flowing in the system since the contacts are assumed to be completely  spin-polarized and anti-parallel.  With the introduction of an all up-spin bath, the electrons from the right contact can enter the dot and interact with the bath causing electrons to spin-flip at the cost of nuclear spin-flops. This spin-flip process may be described  in the form\cite{Basky_Datta} 
$(\downarrow, \Uparrow) \Longleftrightarrow (\uparrow, \Downarrow)$, where $\uparrow, \downarrow$ represent the electronic spin and $\Uparrow, \Downarrow$ represent the nuclear spin. Ordinarily, this ``reaction" would proceed in either direction, which is not the case here owing to a spin-polarized bath.\\
\indent Thus if we start with a case where $F_I = 0$ i.e., $N_\Uparrow = N_\Downarrow$ the bath remains in its equilibrium state, and hence there is no information flow $I_F$.\ But when we begin with $F_I = 1$, the down spin electron entering the dot can interact with the excess up-spin nuclei whereas the up-spin electron entering the dot has no down-spin nucleus to interact with. Thus the resulting up-spin electron can exit from the left contact resulting in current flowing in the system which is opposite to the sense of direction defined in Fig.~\ref{1b}. Now due to the nuclear flops, $F_I$ starts decaying and the increasing $N_\Downarrow$ then allow the inflow of up-spin electron from the left contact, thus opposing the previous current. Consequently current also starts decaying. Now, asymptotically $N_\Uparrow$ approaches $ N_\Downarrow$. Let us say after a long time, $t_0$ we have $N_\Uparrow \approx N_\Downarrow$ and $F_I \approx 0$. The process reaches steady state and there is no current flowing in the dot. The ``recharging of the battery'' is achieved by taking the nuclear bath to its fully polarized state, which is precisely the Landauer's erasure process. As we increase the nuclear spin relaxation constant, the rate at which $F_I$ decays also enhances and so does the rate of decay of the charge current and the information current. These aspects are depicted in Fig.~\ref{2a}, Fig.~\ref{2b} and Fig.~\ref{2c}.\\
\indent We comment on extending this analysis to coherent transient thermal machines. Note that since the two reservoirs are collinear, there are no coherences generated by the evolution. Due to this, we modeled the evolution in terms of rate equations. On the other hand, if the two reservoirs were non-collinear, the evolution would generate coherences in the system \cite{Basky_Milena}. Protocols that extract work from such coherent quantum systems are well known\cite{korzekwa2016extraction} and their implementation in the thermoelectric setup studied here is left for future work. \\
\indent For evaluating the performance of the dot, we make all the quantities dimensionless. We normalize the time with a steady state time constant $t_0 = 1/\gamma_0$, the charge current with $I_0 = 10^3q/t_0$, voltages with $V_0 = 10^{-3} q$, the information current with $I_0^F = k_{B}/t_0$ and the power output with $P_0= V_0I_0$. We have set $\hbar\gamma_0 = 5 \times 10^{-14} eV$ and $\hbar\gamma_{LU} = \hbar\gamma_{RD} = 0.01 meV$.
\subsection{Transient analysis of information-driven battery}
\begin{center}
\begin{figure}[!htb]	
\subfigure[]{\includegraphics[width=0.4\textwidth, height=0.15\textwidth]{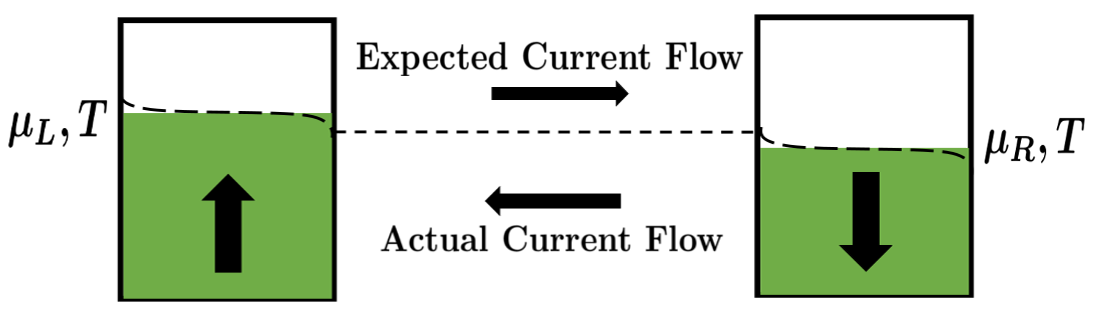}\label{3a}}
\quad	
\subfigure[]{\includegraphics[height=0.221\textwidth,width=0.222\textwidth]{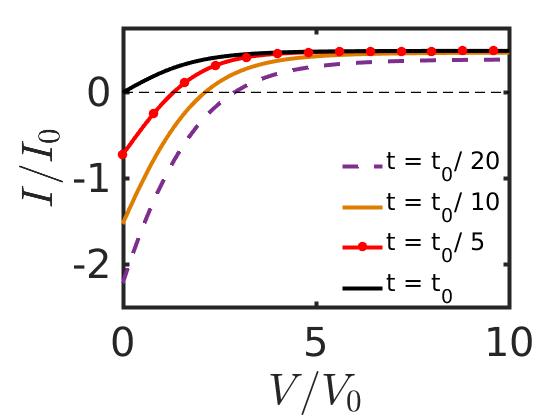}\label{3b}}	
\quad
\subfigure[]{\includegraphics[height=0.221\textwidth,width=0.229\textwidth]{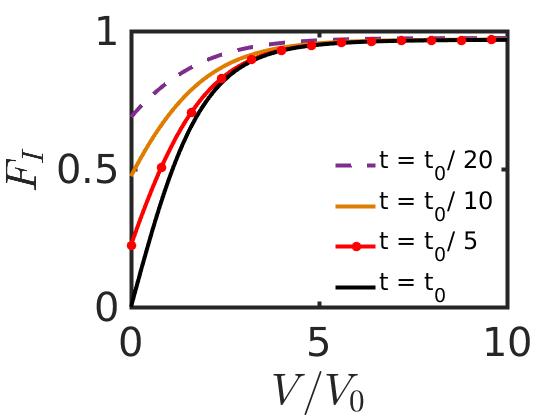}\label{3c}}
\quad
\subfigure[]{\includegraphics[height=0.221\textwidth,width=0.222\textwidth]{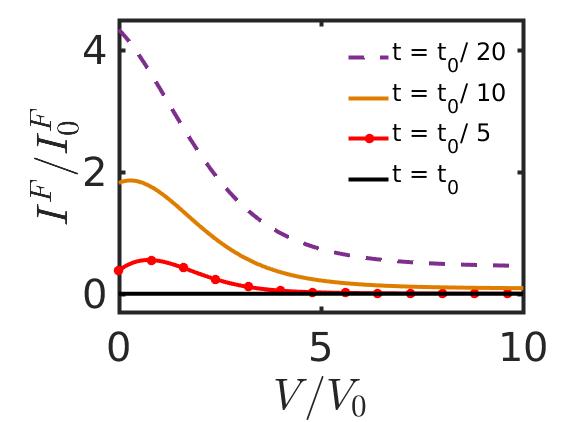}\label{3d}}
\quad
\subfigure[]{\includegraphics[height=0.221\textwidth,width=0.222\textwidth]{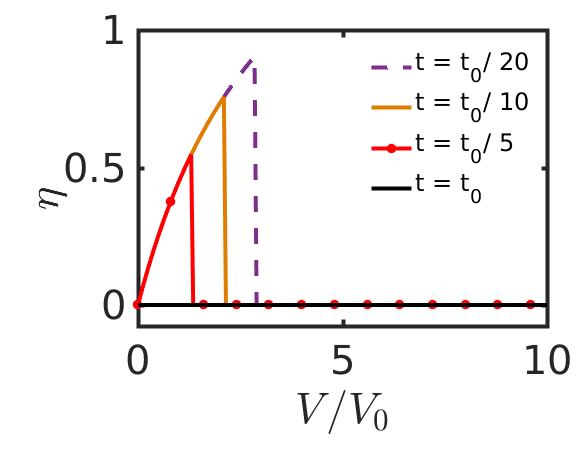}\label{3e}}
\quad
\caption{Transient performance of the information-driven battery with the nuclear relaxation rate set to $\gamma_0$ and $\Delta T = 0$. (a) The system can perform useful work by driving the external current against the terminal voltage. (b) Electronic current-voltage (I-V) characteristics at different time instants. Initially current is dominant due to the presence of only $N_\Uparrow$ in the bath,  whereas the steady state current is dominated by the inflow of up-spin electrons. (c) Variation of $F_I$ as a function of the voltage at different time instants. $F_I$ decays over time with the steady state value depending on the applied voltage bias. (d) Variation of the information current as a function of voltage at different time instants. The information current flowing in the dot results in a finite non-zero current in the dot. (e) Plot of the efficiency as a function of the voltage at various time instants. Due to the negative transient current for a certain voltage range, there is a finite efficiency even when $\Delta T = 0$. }
\end{figure}
\end{center}

We now show that for a range of bias voltage, our device acts as a transient power source. For this, we vary the voltage bias across the contacts keeping their temperatures equal. The left contact being at a higher potential, we would expect a preferential inflow of up-spin electrons, depending on the bias voltage $V=\Delta \mu/q$. But initially we have only $N_\Uparrow$ in the bath, and following the battery operation discussed in the previous section, this will result in a current in the opposite sense as sketched in Fig.~\ref{3a} under transient conditions. The system thus can act as a source of energy driving a load till the voltage bias is equal to the Seebeck voltage because of the non-negative power in this region. Thus there exists a finite thermodynamic efficiency even without a temperature gradient. Initially, we get a good supply of the information current $I_F$ resulting in large transient charge currents.\\
For the nuclear bath to be in steady state, the ``inflow" of spin information into the bath due to the actual charge current flowing through the dot must balance the ``outflow'' of spin information from the bath due to relaxation. Now if we work with $\gamma_I = 0$, then no current flows in the steady state at any voltage bias as there is no outflow of spin information from the bath and hence no information current, $I_F$.  For $\gamma_I = \gamma_0$, the charge current, which balances the outflow of spin information, is due to dominant inflow of up-spin electrons from the right contact and is hence positive as seen in Fig.~\ref{3b}. The steady state nuclear polarization $F_I$ increases with voltage as seen in Fig.~\ref{3c}, since we need an increasing amount of $N_\Uparrow$ to balance increasing inflow of up-spin electrons. In the steady state the bath ``forgets" its positive polarization due to $\gamma_I$ and the positive current results in the inflow of positive polarization. The steady state current thus balances the outflow of spin information from the bath. Due to the decaying $F_I$, $I_F$ also decays over time ultimately becoming zero in the steady state, as seen from Fig.~\ref{3d}. In steady state, the Seebeck voltage and the efficiency also drop to zero as noted in Fig.~\ref{3e}. It is noted here that the thermodynamic efficiency only defines itself in the transient regime, simply due to the counter-intuitive flow of current leading to an extracted power. In Fig.~\ref{3e}, we note that the efficiency is defined in a dynamic sense and has no reference to the related Carnot efficiency since the contact temperatures are equal, and hence Carnot efficiency is technically undefined.
\subsection{Thermoelectrics of information-driven heat engines}
Now we focus on the study of the information-driven heat engine where we also turn on the temperature gradient between the two contacts. We first perform the transient analysis of the heat engine with $\Delta T$ between the contacts corresponding to a Carnot efficiency $\eta_C = 1-T_L/T_R= 0.33$. We also vary the voltage gradient across the contacts in a typical voltage controlled setup discussed in many previous works \cite{Leijnse,Basky_Milena,bitan}.  At low voltages, there is a preferential inflow of down-spin electrons, resulting in a high negative transient current as $F_I$ = 1 initially. The nuclear polarization $F_I$ decays over time with the steady state being negative to oppose the inflow of down-spin electrons. This results in a decay of current. However, a  finite non-zero current at zero voltage bias occurs due to the non-zero $\Delta T$ and non-zero $\gamma_I$. If we had $\gamma_I = 0$, even with a finite $\Delta T$, there would be no current flowing in steady state as there is no outflow of spin information from the bath. As we increase $\gamma_I$, the magnitude of steady state nuclear polarization $F_I$ drops as seen from \eqref{eq:FI}. \\
\begin{center}
\begin{figure}[!htb]
\subfigure[]{\includegraphics[height=0.225\textwidth,width=0.222\textwidth]{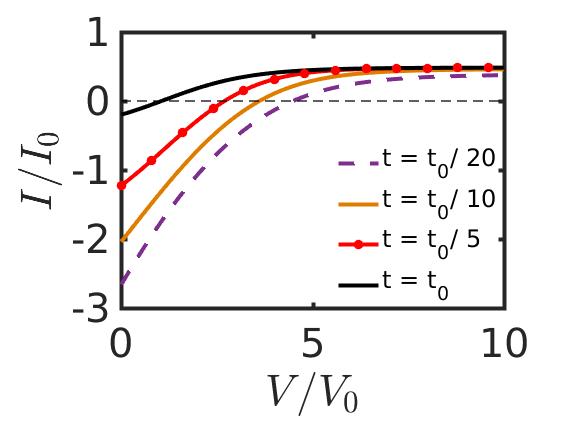}\label{4a}}
\quad
\subfigure[]{\includegraphics[height=0.225\textwidth,width=0.222\textwidth]{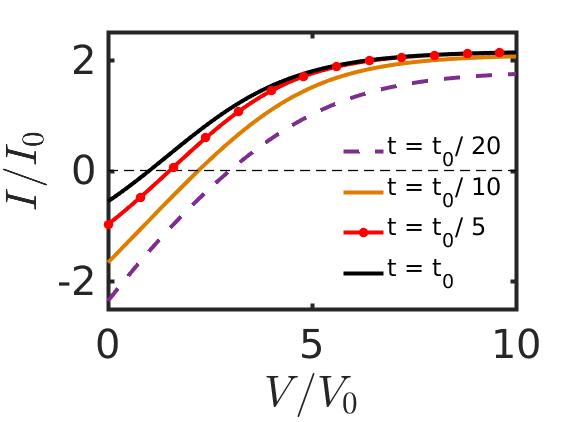}\label{4d}}
\quad
\subfigure[]{\includegraphics[height=0.225\textwidth,width=0.222\textwidth]{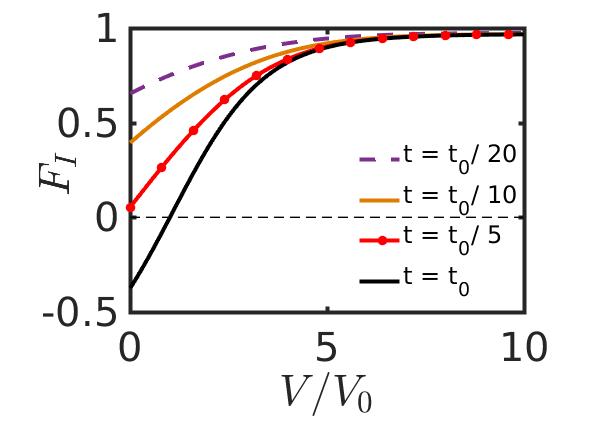}\label{4b}}
\quad
\subfigure[]{\includegraphics[height=0.225\textwidth,width=0.222\textwidth]{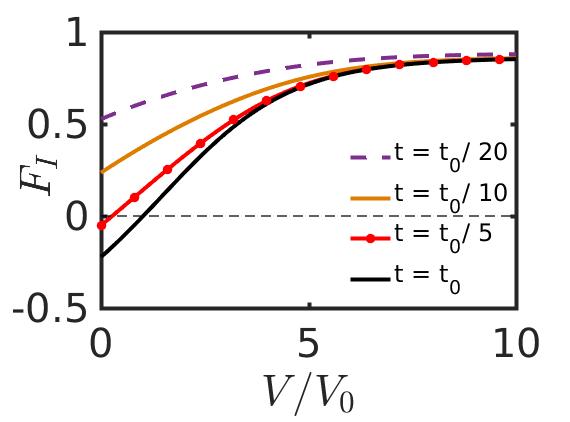}\label{4e}}
\quad
\subfigure[]{\includegraphics[height=0.225\textwidth,width=0.222\textwidth]{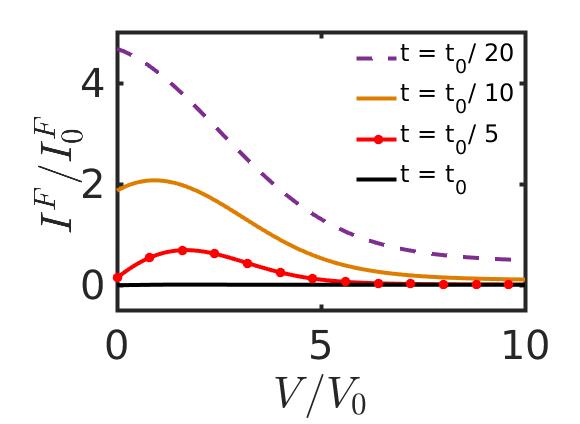}\label{4c}}
\quad
\subfigure[]{\includegraphics[height=0.225\textwidth,width=0.222\textwidth]{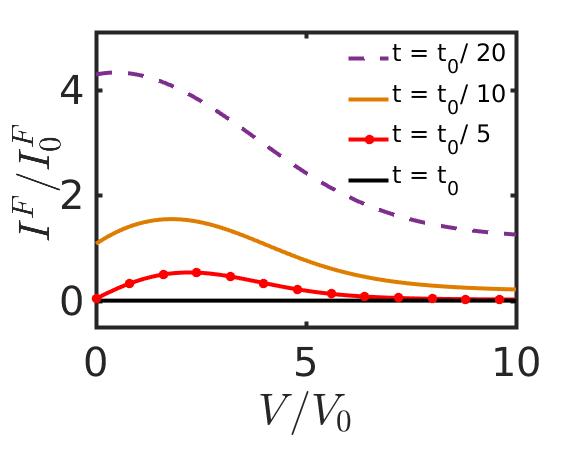}\label{4f}}
\caption{Transient thermoelectric characteristics of the information-driven heat engine with the temperature bias fixed according to the Carnot efficiency set at $\eta_C=0.33$. On the left side we have plots with $\gamma = \gamma_0$ where we compare with the corresponding plots on the right side for which we have $\gamma = 5 \gamma_0$, thus studying the effect of varying nuclear spin relaxation constant. (a),(b) Variation of the charge current as a function of the voltage at different time instants. Current at low voltages is due to $\Delta T$ whereas current at high voltages is due to $\Delta \mu$. (c)(,d) Variation of $F_I$ as a function of the voltage at different time instants. The zero crossing of transient current and the decay of the transient $F_I$ depend on $\gamma_I$. (e),(f) Variation of the information current as a function of the voltage at different time instants. With the decay of $F_I$, the $I_F$ also decays with its magnitude depending $\gamma_I$.}
\end{figure}
\end{center}
\indent For larger values of $\gamma_I$, in steady state there is an increased magnitude of current needed to balance the increased outflow of spin information. At high voltages, there is a preferential inflow of up-spin electrons, and the situation is similar to the transients of the information driven battery discussed previously.  In the steady state, the bath forgets its negative polarization in the low voltage region and the negative current balances this by injecting down spins. Whereas in the high voltage region, the bath forgets its positive polarization which is balanced by a positive current flowing in the dot. We see that in the transient situation it is the information current, which drives a larger magnitude of current but in steady state since the $I_F$ drops down to zero, the current is the same as obtained under steady state conditions using the standard quantum dot heat engine setup\cite{Basky_Milena}.
\begin{center}
\begin{figure}[!htb]
\subfigure[]{\includegraphics[height=0.225\textwidth,width=0.225\textwidth]{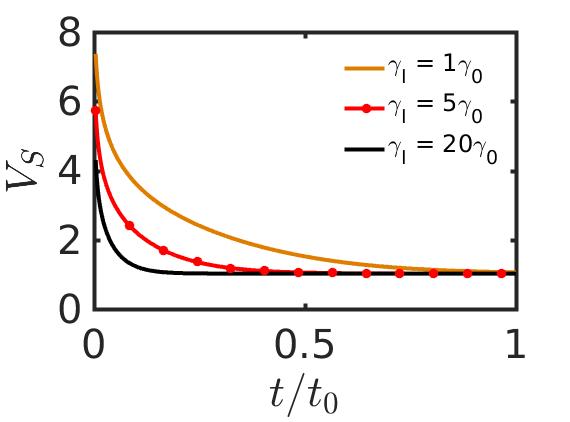}\label{5a}}
\quad
\subfigure[]{\includegraphics[height=0.225\textwidth,width=0.225\textwidth]{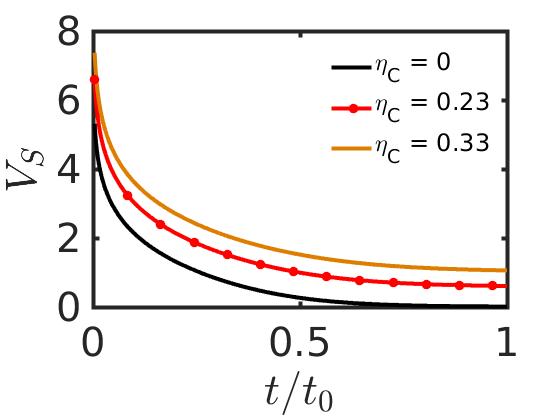}\label{5b}}
\quad
\subfigure[]{\includegraphics[height=0.225\textwidth,width=0.224\textwidth]{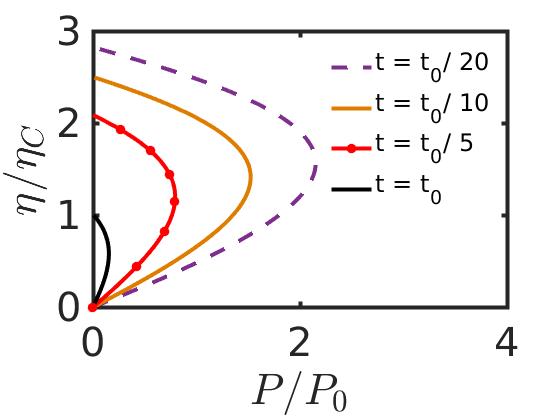}\label{5c}}
\quad
\subfigure[]{\includegraphics[height=0.225\textwidth,width=0.225\textwidth]{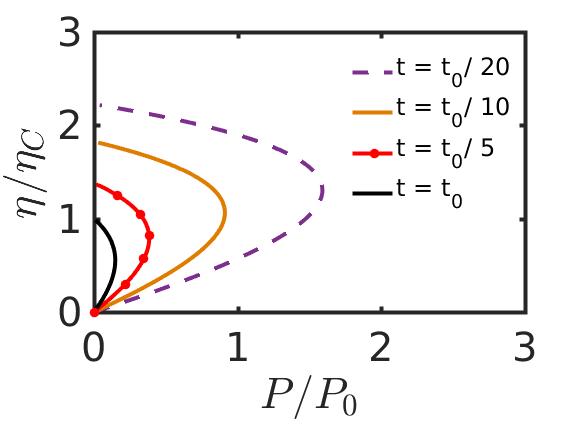}\label{5d}}
\caption{Transient thermoelectric performance of the information-driven heat engine. (a) Variation of Seebeck voltage with time for different relaxation rates with a temperature bias set according to $\eta_C=0.33$. Under transient conditions, $V_S$ varies with $\gamma_I$, but in the steady state, $V_S$ does not depend on $\gamma_I$. (b) Variations in Seebeck voltage with time for different temperature bias across the contacts with $\gamma_I = \gamma_0$. The transient decay is similar in all three cases, but the steady state value depends on the value of $\eta_C$. (c),(d) Plots of power density versus efficiency curves at various time instants. We set $\gamma_I = \gamma_0$ in (c) and compare it with (d) where $\gamma_I = 5\gamma_0$. The efficiency exceeds the Carnot efficiency under transient conditions where we even get a higher power output.}
\end{figure}
\end{center}
\begin{center}
\begin{figure}[!htb]
\subfigure[]{\includegraphics[height=0.221\textwidth,width=0.229\textwidth]{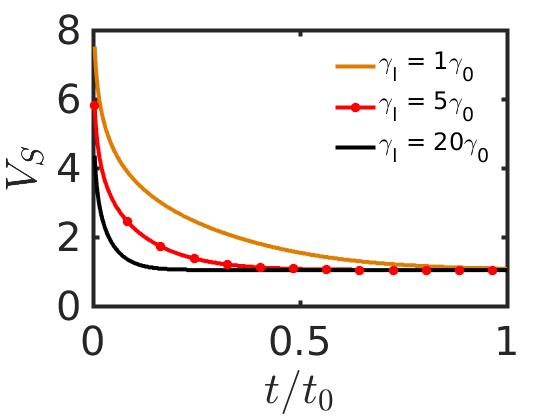}\label{6a}}
\quad
\subfigure[]{\includegraphics[height=0.221\textwidth,width=0.229\textwidth]{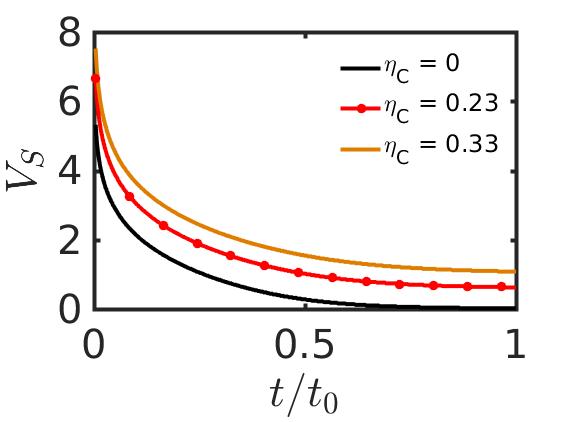}\label{6b}}
\quad
\subfigure[]{\includegraphics[height=0.221\textwidth,width=0.229\textwidth]{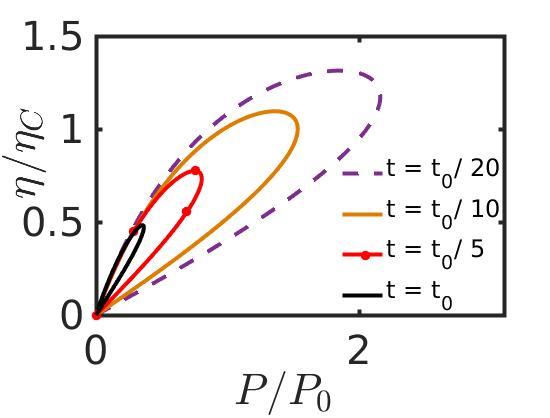}\label{6c}}
\quad
\subfigure[]{\includegraphics[height=0.221\textwidth,width=0.229\textwidth]{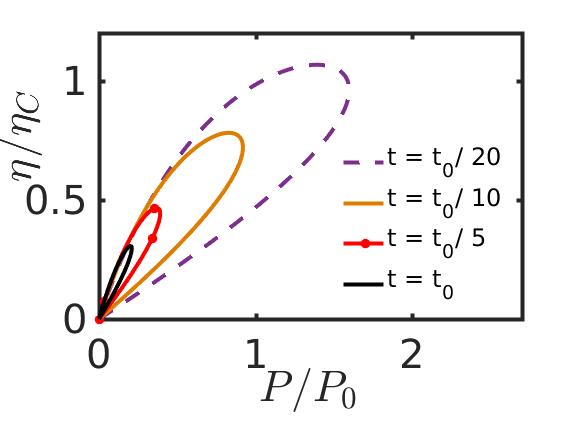}\label{6d}}
\caption{ Transient thermoelectric performance with Coulomb interaction parameter set at $U = k_{B}T_R$ (a) Variations in Seebeck voltage with time for different nuclear relaxation rates with the temperature bias set according to $\eta_C=0.33$. (b) Variations in Seebeck voltage with time for different temperature bias across the contacts with $\gamma_I = \gamma_0$ (c)(d) Plots of efficiency versus power density curves at different times. We set $\gamma_I = \gamma_0$ in (c) and compare it with (d) where $\gamma_I = 5\gamma_0$. }
\end{figure}
\end{center}
\begin{center}
\begin{figure}
\centering
  \includegraphics[height=0.22\textwidth,width=0.35\textwidth]{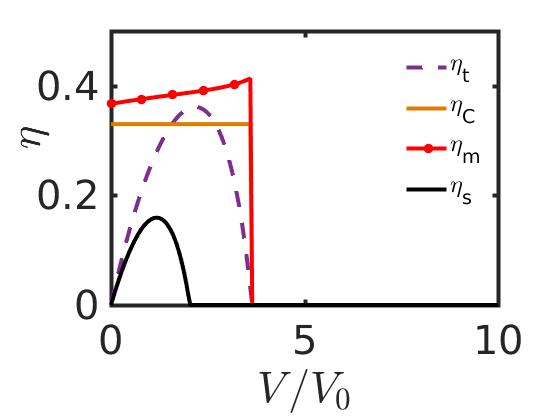}
\caption{ Comparison with the efficiency bound under feedback control. As before we set the Carnot efficiency ($\eta_C = 1-T_L/T_R$) to 0.33. The  Coulomb interaction parameter is set at $U = k_{B}T_R$. The transient efficiency at $t = t_0 / 10$ denoted by $\eta_t$ is always bounded by $\eta_m ( = \eta_C + k_B T_L I_F/I_Q^R)$. At steady state the efficiency ($\eta_s$) drops down and is as expected, bounded by the Carnot efficiency.}\label{fig:Sag}
\end{figure}
\end{center}
\indent At the open circuit voltage or Seebeck voltage $V_S$, we have no information current $I_F$ and hence we also have zero charge current and null nuclear polarization. The Seebeck voltage varies with $\gamma_I$ under transient conditions, but in the steady state, we get the same Seebeck voltage irrespective of the $\gamma_I$ as seen in Fig.~\ref{5a}. However the Seebeck voltage also varies if we vary the temperature bias across the contacts.\ Figure~\ref{5b} depicts this variation in Seeback voltage for different applied temperature bias. Under transient conditions, the voltage for zero crossing of the current and the magnitude of the useful current also depends on $\gamma_I$, resulting in different magnitudes of Seebeck voltage and efficiencies as  $\gamma_I$ is varied. \\
\indent We now introduce the Coulomb interaction parameter\cite{Basky_Grifoni} in the dot.  There is no variation in the Seebeck voltage by varying $\gamma_I$ or by varying $\eta_C$ as noted in Fig.~\ref{6a} and Fig.~\ref{6b}. The variation in heat current due to turning on Coulomb interaction thus affects the efficiency. In the transient  state, we can still exceed the Carnot efficiency due to the information current flow into the dot as noted in Fig.~\ref{6c} and Fig.~\ref{6d}. With the Coulomb interaction, the efficiency does not drop suddenly to zero near the Seebeck voltage as opposed to the case without Coulomb interaction as discussed in Ref. \cite{Basky_Milena}.
\subsection{Efficiency bounds under feedback control}
To understand the efficiency of this thermal machine under feedback control, we focus on Figs.~\ref{5c}, ~\ref{5d}, ~\ref{6c} and ~\ref{6d} . A study of these figures indicates that it is possible to exceed the Carnot efficiency numerically under transient conditions. However, in steady state we perform just as well as the regular quantum dot irrespective of $\gamma_I$ in terms of both the efficiency and Seebeck voltage. This aspect has been hinted upon in a few works that deal with stochastic thermodynamics \cite{Sagawa2008,Seifert2014,Esposito_thermo,Esposito_Demon}. It must, however, be noted that the apparent larger than Carnot operation does not violate the principle behind Carnot's original bound. The efficiency in our case merely exceeds the numerical value of $\eta_C=1-T_L/T_R$, which is the Carnot efficiency taking into account the two temperature baths only.  Since the zero entropy nuclear reservoir can act as a source of extractable work, we should account for this energy when we calculate heat into the system. We must note that our calculation of efficiency from \eqref{eq:eff} does not include this information flow as input in the denominator. If one apparently has a temperature associated with such a Landauer's erasure process, we may be able to craft a Carnot principle which we leave for future work.\\
\indent To cast a firm footing to the above mentioned points, we turn our attention to the nuclear spin bath as the feedback controller \cite{Sagawa2008,Esposito_thermo}. Following \cite{Sagawa2008}, we can draw a parallel with our thermal machine in the following way: a) The electron hops into the quantum dot, b) A measurement is performed by the nuclear bath via a spin-flip process between the electron and nuclear spins. This process allows the electron to transit to the other contact thus completing each cycle. The fact that our spin-flip process is indeed a classical measurement can be shown easily. Under these conditions, it was pointed out that the maximum work extraction from the cycle is related to the mutual information content between the pre-measurement state and the set of outcomes, which in our case is an up-spin or a down-spin. In the dynamic limit \cite{Esposito_thermo}, one replaces the static quantities with the rate of work extraction and rate of heat flow. In this process, the bound now relates to the information current described in this work and becomes $\eta_m ( = \eta_C + k_B T_L I_F/I_Q^R)$. We note from Fig. \ref{fig:Sag}, that the transient efficiency although crosses the Carnot efficiency with this feedback control, it always remains bounded by $\eta_m$.
\section{\label{sec:level4} Conclusion} We investigated a realistic, information driven, quantum dot based quantum thermal machine. The thermal machines we considered include nanoscale batteries and thermoelectric heat engines. The flow of classical information is a source of work, a point well understood from Landauer's erasure. We demonstrated the following fundamental aspects of such a system: (a) The process of Landauer's erasure results can result in a transient nanoscale battery, (b) Even with a voltage bias across the dot, the system can be used as a transient power source to perform useful work, (c) With both the temperature and voltage gradient, the transient thermoelectric response of the dot exceeds the steady state Seebeck voltage and efficiency. We thus conclude that the information current enhances the thermoelectric properties of the system independent of the voltage and temperature bias. Such systems under transient conditions could thus be of use to convert information to useful work over a larger voltage region and to increase the heat to useful work conversion efficiency well in excess of the corresponding Carnot cycle. Our results set the stage for the effective use of scattering processes as a non-equilibrium source of Shannon information flow and the possibilities that may arise from the use of a quantum information source.\\
{\it{Acknowledgments:}} BM acknowledges support from IITB-IRCC grant number 12IRCC013. SV acknowledges support from an IITB-IRCC grant number 16IRCCSG019 and by the National Research Foundation, Prime Minister's Office, Singapore under its Competitive Research Programme (CRP Award No. NRF-CRP14-2014- 02).
\bibliographystyle{apsrev}
\bibliography{references_latest}

\begin{thebibliography}{53}
\expandafter\ifx\csname natexlab\endcsname\relax\def\natexlab#1{#1}\fi
\expandafter\ifx\csname bibnamefont\endcsname\relax
  \def\bibnamefont#1{#1}\fi
\expandafter\ifx\csname bibfnamefont\endcsname\relax
  \def\bibfnamefont#1{#1}\fi
\expandafter\ifx\csname citenamefont\endcsname\relax
  \def\citenamefont#1{#1}\fi
\expandafter\ifx\csname url\endcsname\relax
  \def\url#1{\texttt{#1}}\fi
\expandafter\ifx\csname urlprefix\endcsname\relax\def\urlprefix{URL }\fi
\providecommand{\bibinfo}[2]{#2}
\providecommand{\eprint}[2][]{\url{#2}}

\bibitem[{\citenamefont{Mahan and Sofo}(1996)}]{Mahan1996}
\bibinfo{author}{\bibfnamefont{G.~D.} \bibnamefont{Mahan}} \bibnamefont{and}
  \bibinfo{author}{\bibfnamefont{J.~O.} \bibnamefont{Sofo}},
  \bibinfo{journal}{Proceedings of the National Academy of Sciences of the
  United States of America} \textbf{\bibinfo{volume}{93}},
  \bibinfo{pages}{7436} (\bibinfo{year}{1996}).

\bibitem[{\citenamefont{Andreev and Matveev}(2001)}]{Andreev2001}
\bibinfo{author}{\bibfnamefont{A.~V.} \bibnamefont{Andreev}} \bibnamefont{and}
  \bibinfo{author}{\bibfnamefont{K.~A.} \bibnamefont{Matveev}},
  \bibinfo{journal}{Phys. Rev. Lett} \textbf{\bibinfo{volume}{86}},
  \bibinfo{pages}{280} (\bibinfo{year}{2001}).

\bibitem[{\citenamefont{Humphrey et~al.}(2002)\citenamefont{Humphrey, Newbury,
  Taylor, and Linke}}]{Humphrey2002}
\bibinfo{author}{\bibfnamefont{T.~E.} \bibnamefont{Humphrey}},
  \bibinfo{author}{\bibfnamefont{R.}~\bibnamefont{Newbury}},
  \bibinfo{author}{\bibfnamefont{R.~P.} \bibnamefont{Taylor}},
  \bibnamefont{and} \bibinfo{author}{\bibfnamefont{H.}~\bibnamefont{Linke}},
  \bibinfo{journal}{Phys. Rev. Lett} \textbf{\bibinfo{volume}{89}},
  \bibinfo{pages}{116801} (\bibinfo{year}{2002}).

\bibitem[{\citenamefont{Kubala and K\"{o}nig}(2006)}]{Kubala2006}
\bibinfo{author}{\bibfnamefont{B.}~\bibnamefont{Kubala}} \bibnamefont{and}
  \bibinfo{author}{\bibfnamefont{J.}~\bibnamefont{K\"{o}nig}},
  \bibinfo{journal}{Phys. Rev. B} \textbf{\bibinfo{volume}{73}},
  \bibinfo{pages}{195316} (\bibinfo{year}{2006}).

\bibitem[{\citenamefont{Kubala et~al.}(2008)\citenamefont{Kubala, K\"{o}nig,
  and Pekola}}]{Kubala2008}
\bibinfo{author}{\bibfnamefont{B.}~\bibnamefont{Kubala}},
  \bibinfo{author}{\bibfnamefont{J.}~\bibnamefont{K\"{o}nig}},
  \bibnamefont{and} \bibinfo{author}{\bibfnamefont{J.}~\bibnamefont{Pekola}},
  \bibinfo{journal}{Phys. Rev. Lett} \textbf{\bibinfo{volume}{100}},
  \bibinfo{pages}{066801} (\bibinfo{year}{2008}).

\bibitem[{\citenamefont{Nakpathomkun et~al.}(2010)\citenamefont{Nakpathomkun,
  Xu, and Linke}}]{nak}
\bibinfo{author}{\bibfnamefont{N.}~\bibnamefont{Nakpathomkun}},
  \bibinfo{author}{\bibfnamefont{H.~Q.} \bibnamefont{Xu}}, \bibnamefont{and}
  \bibinfo{author}{\bibfnamefont{H.}~\bibnamefont{Linke}},
  \bibinfo{journal}{Phys. Rev. B} \textbf{\bibinfo{volume}{82}},
  \bibinfo{pages}{235428} (\bibinfo{year}{2010}).

\bibitem[{\citenamefont{Kim et~al.}(2014)\citenamefont{Kim, Jeong, Kim, Lee,
  and Reddy}}]{Kim2014}
\bibinfo{author}{\bibfnamefont{Y.}~\bibnamefont{Kim}},
  \bibinfo{author}{\bibfnamefont{W.}~\bibnamefont{Jeong}},
  \bibinfo{author}{\bibfnamefont{K.}~\bibnamefont{Kim}},
  \bibinfo{author}{\bibfnamefont{W.}~\bibnamefont{Lee}}, \bibnamefont{and}
  \bibinfo{author}{\bibfnamefont{P.}~\bibnamefont{Reddy}},
  \bibinfo{journal}{Nat. Nanotechnol.} \textbf{\bibinfo{volume}{9}},
  \bibinfo{pages}{881} (\bibinfo{year}{2014}).

\bibitem[{\citenamefont{Reddy et~al.}(2007)\citenamefont{Reddy, Jang, Segalman,
  and Majumdar}}]{Reddy2007}
\bibinfo{author}{\bibfnamefont{P.}~\bibnamefont{Reddy}},
  \bibinfo{author}{\bibfnamefont{S.-Y.} \bibnamefont{Jang}},
  \bibinfo{author}{\bibfnamefont{R.~A.} \bibnamefont{Segalman}},
  \bibnamefont{and} \bibinfo{author}{\bibfnamefont{A.}~\bibnamefont{Majumdar}},
  \bibinfo{journal}{Science} \textbf{\bibinfo{volume}{315}},
  \bibinfo{pages}{1568} (\bibinfo{year}{2007}).

\bibitem[{\citenamefont{Sothmann et~al.}(2013)\citenamefont{Sothmann, Sanchez,
  Jordan, and Buttiker}}]{jordan1}
\bibinfo{author}{\bibfnamefont{B.}~\bibnamefont{Sothmann}},
  \bibinfo{author}{\bibfnamefont{R.}~\bibnamefont{Sanchez}},
  \bibinfo{author}{\bibfnamefont{A.}~\bibnamefont{Jordan}}, \bibnamefont{and}
  \bibinfo{author}{\bibfnamefont{M.}~\bibnamefont{Buttiker}},
  \bibinfo{journal}{New J. Phys.} \textbf{\bibinfo{volume}{15}},
  \bibinfo{pages}{095021} (\bibinfo{year}{2013}).

\bibitem[{\citenamefont{Choi and Jordan}(2016)}]{jordan2}
\bibinfo{author}{\bibfnamefont{Y.}~\bibnamefont{Choi}} \bibnamefont{and}
  \bibinfo{author}{\bibfnamefont{A.~N.} \bibnamefont{Jordan}},
  \bibinfo{journal}{Physica E} \textbf{\bibinfo{volume}{74}},
  \bibinfo{pages}{465} (\bibinfo{year}{2016}).

\bibitem[{\citenamefont{Agarwal and Muralidharan}(2014)}]{Agarwal2014}
\bibinfo{author}{\bibfnamefont{A.}~\bibnamefont{Agarwal}} \bibnamefont{and}
  \bibinfo{author}{\bibfnamefont{B.}~\bibnamefont{Muralidharan}},
  \bibinfo{journal}{App. Phys. Lett.} \textbf{\bibinfo{volume}{105}},
  \bibinfo{pages}{013104} (\bibinfo{year}{2014}).

\bibitem[{\citenamefont{{Whitney}}(2014)}]{whitney}
\bibinfo{author}{\bibfnamefont{R.~S.} \bibnamefont{{Whitney}}},
  \bibinfo{journal}{Physical Review Letters} \textbf{\bibinfo{volume}{112}},
  \bibinfo{eid}{130601} (\bibinfo{year}{2014}).

\bibitem[{\citenamefont{Muralidharan and Grifoni}(2012)}]{Basky_Grifoni}
\bibinfo{author}{\bibfnamefont{B.}~\bibnamefont{Muralidharan}}
  \bibnamefont{and} \bibinfo{author}{\bibfnamefont{M.}~\bibnamefont{Grifoni}},
  \bibinfo{journal}{Phys. Rev. B} \textbf{\bibinfo{volume}{85}},
  \bibinfo{pages}{155423} (\bibinfo{year}{2012}).

\bibitem[{\citenamefont{Zimbovskaya}(2016)}]{Zimb2016}
\bibinfo{author}{\bibfnamefont{N.~A.} \bibnamefont{Zimbovskaya}},
  \bibinfo{journal}{Journal of Physics: Condensed Matter}
  \textbf{\bibinfo{volume}{28}}, \bibinfo{pages}{183002}
  (\bibinfo{year}{2016}).

\bibitem[{\citenamefont{Leijnse et~al.}(2010)\citenamefont{Leijnse, Wegewijs,
  and Flensberg}}]{Leijnse}
\bibinfo{author}{\bibfnamefont{M.}~\bibnamefont{Leijnse}},
  \bibinfo{author}{\bibfnamefont{M.~R.} \bibnamefont{Wegewijs}},
  \bibnamefont{and}
  \bibinfo{author}{\bibfnamefont{K.}~\bibnamefont{Flensberg}},
  \bibinfo{journal}{Phys. Rev. B} \textbf{\bibinfo{volume}{82}},
  \bibinfo{pages}{045412} (\bibinfo{year}{2010}).

\bibitem[{\citenamefont{Muralidharan and Grifoni}(2013)}]{Basky_Milena}
\bibinfo{author}{\bibfnamefont{B.}~\bibnamefont{Muralidharan}}
  \bibnamefont{and} \bibinfo{author}{\bibfnamefont{M.}~\bibnamefont{Grifoni}},
  \bibinfo{journal}{Phys. Rev. B} \textbf{\bibinfo{volume}{88}},
  \bibinfo{pages}{045402} (\bibinfo{year}{2013}).

\bibitem[{\citenamefont{Jordan et~al.}(2013)\citenamefont{Jordan, Sothmann,
  S\'anchez, and B\"uttiker}}]{jordan3}
\bibinfo{author}{\bibfnamefont{A.~N.} \bibnamefont{Jordan}},
  \bibinfo{author}{\bibfnamefont{B.}~\bibnamefont{Sothmann}},
  \bibinfo{author}{\bibfnamefont{R.}~\bibnamefont{S\'anchez}},
  \bibnamefont{and}
  \bibinfo{author}{\bibfnamefont{M.}~\bibnamefont{B\"uttiker}},
  \bibinfo{journal}{Phys. Rev. B} \textbf{\bibinfo{volume}{87}},
  \bibinfo{pages}{075312} (\bibinfo{year}{2013}).

\bibitem[{\citenamefont{Sothmann et~al.}(2015)\citenamefont{Sothmann,
  S{\'a}nchez, and Jordan}}]{sothmann}
\bibinfo{author}{\bibfnamefont{B.}~\bibnamefont{Sothmann}},
  \bibinfo{author}{\bibfnamefont{R.}~\bibnamefont{S{\'a}nchez}},
  \bibnamefont{and} \bibinfo{author}{\bibfnamefont{A.~N.}
  \bibnamefont{Jordan}}, \bibinfo{journal}{Nanotechnology}
  \textbf{\bibinfo{volume}{26}}, \bibinfo{pages}{032001}
  (\bibinfo{year}{2015}).

\bibitem[{\citenamefont{De and Muralidharan}(2016)}]{bitan}
\bibinfo{author}{\bibfnamefont{B.}~\bibnamefont{De}} \bibnamefont{and}
  \bibinfo{author}{\bibfnamefont{B.}~\bibnamefont{Muralidharan}},
  \bibinfo{journal}{Phys. Rev. B} \textbf{\bibinfo{volume}{94}},
  \bibinfo{pages}{165416} (\bibinfo{year}{2016}).

\bibitem[{\citenamefont{S\'anchez and B\"uttiker}(2011)}]{Sanchez}
\bibinfo{author}{\bibfnamefont{R.}~\bibnamefont{S\'anchez}} \bibnamefont{and}
  \bibinfo{author}{\bibfnamefont{M.}~\bibnamefont{B\"uttiker}},
  \bibinfo{journal}{Phys. Rev. B} \textbf{\bibinfo{volume}{83}},
  \bibinfo{pages}{085428} (\bibinfo{year}{2011}).

\bibitem[{\citenamefont{{Datta}}(2007)}]{Datta_Demon}
\bibinfo{author}{\bibfnamefont{S.}~\bibnamefont{{Datta}}},
  \bibinfo{journal}{ArXiv e-prints}  (\bibinfo{year}{2007}),
  \eprint{0704.1623}.

\bibitem[{\citenamefont{Abreu and Seifert}(2011)}]{Abreu_Seifert}
\bibinfo{author}{\bibfnamefont{D.}~\bibnamefont{Abreu}} \bibnamefont{and}
  \bibinfo{author}{\bibfnamefont{U.}~\bibnamefont{Seifert}},
  \bibinfo{journal}{EPL (Europhysics Letters)} \textbf{\bibinfo{volume}{94}},
  \bibinfo{pages}{10001} (\bibinfo{year}{2011}).

\bibitem[{\citenamefont{Bauer et~al.}(2012)\citenamefont{Bauer, Abreu, and
  Seifert}}]{Bauer_Seifert}
\bibinfo{author}{\bibfnamefont{M.}~\bibnamefont{Bauer}},
  \bibinfo{author}{\bibfnamefont{D.}~\bibnamefont{Abreu}}, \bibnamefont{and}
  \bibinfo{author}{\bibfnamefont{U.}~\bibnamefont{Seifert}},
  \bibinfo{journal}{Journal of Physics A: Mathematical and Theoretical}
  \textbf{\bibinfo{volume}{45}}, \bibinfo{pages}{162001}
  (\bibinfo{year}{2012}).

\bibitem[{\citenamefont{Abreu and Seifert}(2012)}]{David_Seifert}
\bibinfo{author}{\bibfnamefont{D.}~\bibnamefont{Abreu}} \bibnamefont{and}
  \bibinfo{author}{\bibfnamefont{U.}~\bibnamefont{Seifert}},
  \bibinfo{journal}{Phys. Rev. Lett.} \textbf{\bibinfo{volume}{108}},
  \bibinfo{pages}{030601} (\bibinfo{year}{2012}).

\bibitem[{\citenamefont{Barato and Seifert}(2014)}]{Seifert2014}
\bibinfo{author}{\bibfnamefont{A.~C.} \bibnamefont{Barato}} \bibnamefont{and}
  \bibinfo{author}{\bibfnamefont{U.}~\bibnamefont{Seifert}},
  \bibinfo{journal}{Phys. Rev. E} \textbf{\bibinfo{volume}{90}},
  \bibinfo{pages}{042150} (\bibinfo{year}{2014}).

\bibitem[{\citenamefont{Esposito and Schaller}(2012)}]{Esposito_thermo}
\bibinfo{author}{\bibfnamefont{M.}~\bibnamefont{Esposito}} \bibnamefont{and}
  \bibinfo{author}{\bibfnamefont{G.}~\bibnamefont{Schaller}},
  \bibinfo{journal}{EPL (Europhysics Letters)} \textbf{\bibinfo{volume}{99}},
  \bibinfo{pages}{30003} (\bibinfo{year}{2012}).

\bibitem[{\citenamefont{Strasberg et~al.}(2013)\citenamefont{Strasberg,
  Schaller, Brandes, and Esposito}}]{Esposito_Demon}
\bibinfo{author}{\bibfnamefont{P.}~\bibnamefont{Strasberg}},
  \bibinfo{author}{\bibfnamefont{G.}~\bibnamefont{Schaller}},
  \bibinfo{author}{\bibfnamefont{T.}~\bibnamefont{Brandes}}, \bibnamefont{and}
  \bibinfo{author}{\bibfnamefont{M.}~\bibnamefont{Esposito}},
  \bibinfo{journal}{Phys. Rev. Lett.} \textbf{\bibinfo{volume}{110}},
  \bibinfo{pages}{040601} (\bibinfo{year}{2013}).

\bibitem[{\citenamefont{Averin and Pekola}(2017)}]{Averin}
\bibinfo{author}{\bibfnamefont{D.~V.} \bibnamefont{Averin}} \bibnamefont{and}
  \bibinfo{author}{\bibfnamefont{J.~P.} \bibnamefont{Pekola}},
  \bibinfo{journal}{physica status solidi (b)} \textbf{\bibinfo{volume}{254}},
  \bibinfo{pages}{1600677} (\bibinfo{year}{2017}), ISSN
  \bibinfo{issn}{1521-3951}.

\bibitem[{\citenamefont{Strasberg et~al.}(2014)\citenamefont{Strasberg,
  Schaller, Brandes, and Jarzynski}}]{Jarzynski}
\bibinfo{author}{\bibfnamefont{P.}~\bibnamefont{Strasberg}},
  \bibinfo{author}{\bibfnamefont{G.}~\bibnamefont{Schaller}},
  \bibinfo{author}{\bibfnamefont{T.}~\bibnamefont{Brandes}}, \bibnamefont{and}
  \bibinfo{author}{\bibfnamefont{C.}~\bibnamefont{Jarzynski}},
  \bibinfo{journal}{Phys. Rev. E} \textbf{\bibinfo{volume}{90}},
  \bibinfo{pages}{062107} (\bibinfo{year}{2014}).

\bibitem[{\citenamefont{W\"urtz et~al.}(2005)\citenamefont{W\"urtz, M\"uller,
  Lorke, Reuter, and Wieck}}]{Wieck}
\bibinfo{author}{\bibfnamefont{A.}~\bibnamefont{W\"urtz}},
  \bibinfo{author}{\bibfnamefont{T.}~\bibnamefont{M\"uller}},
  \bibinfo{author}{\bibfnamefont{A.}~\bibnamefont{Lorke}},
  \bibinfo{author}{\bibfnamefont{D.}~\bibnamefont{Reuter}}, \bibnamefont{and}
  \bibinfo{author}{\bibfnamefont{A.~D.} \bibnamefont{Wieck}},
  \bibinfo{journal}{Phys. Rev. Lett.} \textbf{\bibinfo{volume}{95}},
  \bibinfo{pages}{056802} (\bibinfo{year}{2005}).

\bibitem[{\citenamefont{Sagawa and Ueda}(2008)}]{Sagawa2008}
\bibinfo{author}{\bibfnamefont{T.}~\bibnamefont{Sagawa}} \bibnamefont{and}
  \bibinfo{author}{\bibfnamefont{M.}~\bibnamefont{Ueda}},
  \bibinfo{journal}{Phys. Rev. Lett.} \textbf{\bibinfo{volume}{100}},
  \bibinfo{pages}{080403} (\bibinfo{year}{2008}).

\bibitem[{\citenamefont{Sagawa and Ueda}(2009)}]{Sagawa2009}
\bibinfo{author}{\bibfnamefont{T.}~\bibnamefont{Sagawa}} \bibnamefont{and}
  \bibinfo{author}{\bibfnamefont{M.}~\bibnamefont{Ueda}},
  \bibinfo{journal}{Phys. Rev. Lett.} \textbf{\bibinfo{volume}{102}},
  \bibinfo{pages}{250602} (\bibinfo{year}{2009}).

\bibitem[{\citenamefont{Sagawa and Ueda}(2010)}]{Sagawa2010}
\bibinfo{author}{\bibfnamefont{T.}~\bibnamefont{Sagawa}} \bibnamefont{and}
  \bibinfo{author}{\bibfnamefont{M.}~\bibnamefont{Ueda}},
  \bibinfo{journal}{Phys. Rev. Lett.} \textbf{\bibinfo{volume}{104}},
  \bibinfo{pages}{090602} (\bibinfo{year}{2010}).

\bibitem[{\citenamefont{Koski et~al.}(2014)\citenamefont{Koski, Maisi, Sagawa,
  and Pekola}}]{Sagawa2014}
\bibinfo{author}{\bibfnamefont{J.~V.} \bibnamefont{Koski}},
  \bibinfo{author}{\bibfnamefont{V.~F.} \bibnamefont{Maisi}},
  \bibinfo{author}{\bibfnamefont{T.}~\bibnamefont{Sagawa}}, \bibnamefont{and}
  \bibinfo{author}{\bibfnamefont{J.~P.} \bibnamefont{Pekola}},
  \bibinfo{journal}{Phys. Rev. Lett.} \textbf{\bibinfo{volume}{113}},
  \bibinfo{pages}{030601} (\bibinfo{year}{2014}).

\bibitem[{\citenamefont{Goold et~al.}(2016)\citenamefont{Goold, Huber, Riera,
  del Rio, and Skrzypczyk}}]{goold2016role}
\bibinfo{author}{\bibfnamefont{J.}~\bibnamefont{Goold}},
  \bibinfo{author}{\bibfnamefont{M.}~\bibnamefont{Huber}},
  \bibinfo{author}{\bibfnamefont{A.}~\bibnamefont{Riera}},
  \bibinfo{author}{\bibfnamefont{L.}~\bibnamefont{del Rio}}, \bibnamefont{and}
  \bibinfo{author}{\bibfnamefont{P.}~\bibnamefont{Skrzypczyk}},
  \bibinfo{journal}{Journal of Physics A: Mathematical and Theoretical}
  \textbf{\bibinfo{volume}{49}}, \bibinfo{pages}{143001}
  (\bibinfo{year}{2016}).

\bibitem[{\citenamefont{Vinjanampathy and
  Anders}(2016)}]{vinjanampathy2016quantum}
\bibinfo{author}{\bibfnamefont{S.}~\bibnamefont{Vinjanampathy}}
  \bibnamefont{and} \bibinfo{author}{\bibfnamefont{J.}~\bibnamefont{Anders}},
  \bibinfo{journal}{Contemporary Physics} \textbf{\bibinfo{volume}{57}},
  \bibinfo{pages}{545} (\bibinfo{year}{2016}).

\bibitem[{\citenamefont{Hanson et~al.}(2007)\citenamefont{Hanson, Kouwenhoven,
  Petta, Tarucha, and Vandersypen}}]{Hanson2007}
\bibinfo{author}{\bibfnamefont{R.}~\bibnamefont{Hanson}},
  \bibinfo{author}{\bibfnamefont{L.~P.} \bibnamefont{Kouwenhoven}},
  \bibinfo{author}{\bibfnamefont{J.~R.} \bibnamefont{Petta}},
  \bibinfo{author}{\bibfnamefont{S.}~\bibnamefont{Tarucha}}, \bibnamefont{and}
  \bibinfo{author}{\bibfnamefont{L.~M.~K.} \bibnamefont{Vandersypen}},
  \bibinfo{journal}{Rev. Mod. Phys.} \textbf{\bibinfo{volume}{79}},
  \bibinfo{pages}{1217} (\bibinfo{year}{2007}).

\bibitem[{\citenamefont{Rudner and Levitov}(2007)}]{Levitov1}
\bibinfo{author}{\bibfnamefont{M.~S.} \bibnamefont{Rudner}} \bibnamefont{and}
  \bibinfo{author}{\bibfnamefont{L.~S.} \bibnamefont{Levitov}},
  \bibinfo{journal}{Phys. Rev. Lett.} \textbf{\bibinfo{volume}{99}},
  \bibinfo{pages}{036602} (\bibinfo{year}{2007}).

\bibitem[{\citenamefont{Rudner et~al.}(2011)\citenamefont{Rudner, Koppens,
  Folk, Vandersypen, and Levitov}}]{Levitov2}
\bibinfo{author}{\bibfnamefont{M.~S.} \bibnamefont{Rudner}},
  \bibinfo{author}{\bibfnamefont{F.~H.~L.} \bibnamefont{Koppens}},
  \bibinfo{author}{\bibfnamefont{J.~A.} \bibnamefont{Folk}},
  \bibinfo{author}{\bibfnamefont{L.~M.~K.} \bibnamefont{Vandersypen}},
  \bibnamefont{and} \bibinfo{author}{\bibfnamefont{L.~S.}
  \bibnamefont{Levitov}}, \bibinfo{journal}{Phys. Rev. B}
  \textbf{\bibinfo{volume}{84}}, \bibinfo{pages}{075339}
  (\bibinfo{year}{2011}).

\bibitem[{\citenamefont{Rudner and Levitov}(2010)}]{Levitov3}
\bibinfo{author}{\bibfnamefont{M.~S.} \bibnamefont{Rudner}} \bibnamefont{and}
  \bibinfo{author}{\bibfnamefont{L.~S.} \bibnamefont{Levitov}},
  \bibinfo{journal}{Nanotechnology} \textbf{\bibinfo{volume}{21}},
  \bibinfo{pages}{274016} (\bibinfo{year}{2010}).

\bibitem[{\citenamefont{Buddhiraju and Muralidharan}(2014)}]{Basky_buddhi}
\bibinfo{author}{\bibfnamefont{S.}~\bibnamefont{Buddhiraju}} \bibnamefont{and}
  \bibinfo{author}{\bibfnamefont{B.}~\bibnamefont{Muralidharan}},
  \bibinfo{journal}{Journal of Physics: Condensed Matter}
  \textbf{\bibinfo{volume}{26}}, \bibinfo{pages}{485302}
  (\bibinfo{year}{2014}).

\bibitem[{\citenamefont{Singha et~al.}(2017)\citenamefont{Singha, Fauzi,
  Hirayama, and Muralidharan}}]{Aniket}
\bibinfo{author}{\bibfnamefont{A.}~\bibnamefont{Singha}},
  \bibinfo{author}{\bibfnamefont{M.~H.} \bibnamefont{Fauzi}},
  \bibinfo{author}{\bibfnamefont{Y.}~\bibnamefont{Hirayama}}, \bibnamefont{and}
  \bibinfo{author}{\bibfnamefont{B.}~\bibnamefont{Muralidharan}},
  \bibinfo{journal}{Phys. Rev. B} \textbf{\bibinfo{volume}{95}},
  \bibinfo{pages}{115416} (\bibinfo{year}{2017}).

\bibitem[{\citenamefont{I\~narrea
  et~al.}(2007{\natexlab{a}})\citenamefont{I\~narrea, Platero, and
  MacDonald}}]{Jesus}
\bibinfo{author}{\bibfnamefont{J.}~\bibnamefont{I\~narrea}},
  \bibinfo{author}{\bibfnamefont{G.}~\bibnamefont{Platero}}, \bibnamefont{and}
  \bibinfo{author}{\bibfnamefont{A.~H.} \bibnamefont{MacDonald}},
  \bibinfo{journal}{Phys. Rev. B} \textbf{\bibinfo{volume}{76}},
  \bibinfo{pages}{085329} (\bibinfo{year}{2007}{\natexlab{a}}).

\bibitem[{\citenamefont{Beenakker}(1991)}]{Beenakker}
\bibinfo{author}{\bibfnamefont{C.~W.~J.} \bibnamefont{Beenakker}},
  \bibinfo{journal}{Phys. Rev. B} \textbf{\bibinfo{volume}{44}},
  \bibinfo{pages}{1646} (\bibinfo{year}{1991}).

\bibitem[{\citenamefont{Muralidharan et~al.}(2006)\citenamefont{Muralidharan,
  Ghosh, and Datta}}]{Basky_Ghosh}
\bibinfo{author}{\bibfnamefont{B.}~\bibnamefont{Muralidharan}},
  \bibinfo{author}{\bibfnamefont{A.~W.} \bibnamefont{Ghosh}}, \bibnamefont{and}
  \bibinfo{author}{\bibfnamefont{S.}~\bibnamefont{Datta}},
  \bibinfo{journal}{Phys. Rev. B} \textbf{\bibinfo{volume}{73}},
  \bibinfo{pages}{155410} (\bibinfo{year}{2006}).

\bibitem[{\citenamefont{Muralidharan and Datta}(2007)}]{Basky_Datta}
\bibinfo{author}{\bibfnamefont{B.}~\bibnamefont{Muralidharan}}
  \bibnamefont{and} \bibinfo{author}{\bibfnamefont{S.}~\bibnamefont{Datta}},
  \bibinfo{journal}{Phys. Rev. B} \textbf{\bibinfo{volume}{76}},
  \bibinfo{pages}{035432} (\bibinfo{year}{2007}).

\bibitem[{\citenamefont{Timm}(2008)}]{Timm}
\bibinfo{author}{\bibfnamefont{C.}~\bibnamefont{Timm}}, \bibinfo{journal}{Phys.
  Rev. B} \textbf{\bibinfo{volume}{77}}, \bibinfo{pages}{195416}
  (\bibinfo{year}{2008}).

\bibitem[{\citenamefont{K{\"o}nig and Martinek}(2003)}]{Koenig_1}
\bibinfo{author}{\bibfnamefont{J.}~\bibnamefont{K{\"o}nig}} \bibnamefont{and}
  \bibinfo{author}{\bibfnamefont{J.}~\bibnamefont{Martinek}},
  \bibinfo{journal}{Phys. Rev. Lett.} \textbf{\bibinfo{volume}{90}},
  \bibinfo{pages}{166602} (\bibinfo{year}{2003}).

\bibitem[{\citenamefont{Braun et~al.}(2004)\citenamefont{Braun, K{\"o}nig, and
  Martinek}}]{Koenig_2}
\bibinfo{author}{\bibfnamefont{M.}~\bibnamefont{Braun}},
  \bibinfo{author}{\bibfnamefont{J.}~\bibnamefont{K{\"o}nig}},
  \bibnamefont{and} \bibinfo{author}{\bibfnamefont{J.}~\bibnamefont{Martinek}},
  \bibinfo{journal}{Phys. Rev. B} \textbf{\bibinfo{volume}{70}},
  \bibinfo{pages}{195345} (\bibinfo{year}{2004}).

\bibitem[{\citenamefont{Braig and Brouwer}(2005)}]{Brouw}
\bibinfo{author}{\bibfnamefont{S.}~\bibnamefont{Braig}} \bibnamefont{and}
  \bibinfo{author}{\bibfnamefont{P.~W.} \bibnamefont{Brouwer}},
  \bibinfo{journal}{Phys. Rev. B} \textbf{\bibinfo{volume}{71}},
  \bibinfo{pages}{195324} (\bibinfo{year}{2005}).

\bibitem[{\citenamefont{Hornberger et~al.}(2008)\citenamefont{Hornberger,
  Koller, Begemann, Donarini, and Grifoni}}]{Milena_noncoll}
\bibinfo{author}{\bibfnamefont{R.}~\bibnamefont{Hornberger}},
  \bibinfo{author}{\bibfnamefont{S.}~\bibnamefont{Koller}},
  \bibinfo{author}{\bibfnamefont{G.}~\bibnamefont{Begemann}},
  \bibinfo{author}{\bibfnamefont{A.}~\bibnamefont{Donarini}}, \bibnamefont{and}
  \bibinfo{author}{\bibfnamefont{M.}~\bibnamefont{Grifoni}},
  \bibinfo{journal}{Phys. Rev. B} \textbf{\bibinfo{volume}{77}},
  \bibinfo{pages}{245313} (\bibinfo{year}{2008}).

\bibitem[{\citenamefont{I\~narrea
  et~al.}(2007{\natexlab{b}})\citenamefont{I\~narrea, Platero, and
  MacDonald}}]{MacDonald}
\bibinfo{author}{\bibfnamefont{J.}~\bibnamefont{I\~narrea}},
  \bibinfo{author}{\bibfnamefont{G.}~\bibnamefont{Platero}}, \bibnamefont{and}
  \bibinfo{author}{\bibfnamefont{A.~H.} \bibnamefont{MacDonald}},
  \bibinfo{journal}{Phys. Rev. B} \textbf{\bibinfo{volume}{76}},
  \bibinfo{pages}{085329} (\bibinfo{year}{2007}{\natexlab{b}}).

\bibitem[{\citenamefont{Korzekwa et~al.}(2016)\citenamefont{Korzekwa,
  Lostaglio, Oppenheim, and Jennings}}]{korzekwa2016extraction}
\bibinfo{author}{\bibfnamefont{K.}~\bibnamefont{Korzekwa}},
  \bibinfo{author}{\bibfnamefont{M.}~\bibnamefont{Lostaglio}},
  \bibinfo{author}{\bibfnamefont{J.}~\bibnamefont{Oppenheim}},
  \bibnamefont{and} \bibinfo{author}{\bibfnamefont{D.}~\bibnamefont{Jennings}},
  \bibinfo{journal}{New Journal of Physics} \textbf{\bibinfo{volume}{18}},
  \bibinfo{pages}{023045} (\bibinfo{year}{2016}).

\end{thebibliography}
\end{document}